\documentclass[12pt]{article}

\usepackage{epsfig}
\usepackage{latexsym}

\begin{document}

\noindent
DESY 02-088\hfill{\tt hep-lat/0206022}\\
June 2002
\bigskip
\begin{center}
{\LARGE\bf
Modified iterative versus Laplacian Landau gauge in compact U(1) theory}
\end{center}

\begin{center}
{\large\bf Stephan D\"urr$\,{}^{a,b,c}$}
{\large and}
{\large\bf Philippe de Forcrand$\,{}^{d,e}$}
\vspace{5pt}
\\
${}^a${\small\it DESY Zeuthen, 15738 Zeuthen, Germany}\\
${}^b${\small\it INT at University of Washington, Seattle WA 98195-1550, USA}\\
${}^c${\small\it PSI Villigen, 5232 Villigen PSI, Switzerland}\\
${}^d${\small\it ETH Z\"urich-H\"onggerberg, 8093 Z\"urich, Switzerland}\\
${}^e${\small\it CERN, Theory Division, 1211 Geneva 23, Switzerland}\\
\end{center}

\begin{abstract}
\noindent
Compact U(1) theory in 4 dimensions is used to compare the modified
iterative and the Laplacian fixing to lattice Landau gauge in a controlled
setting, since in the Coulomb phase the lattice theory must reproduce the
perturbative prediction.
It turns out that on either side of the phase transition clear differences show
up and in the Coulomb phase the ability to remove double Dirac sheets proves
vital on a small lattice.
\end{abstract}


\hyphenation{
simu-la-tion theo-re-ti-cal modi-fied
maxi-mum maxi-mize maxi-mizing 
mini-mum mini-mize mini-mizing 
}


\newcommand{\pa}{\partial}
\newcommand{\pas}{\partial\!\!\!/}
\newcommand{\Dsl}{D\!\!\!\!/\,}
\newcommand{\Psl}{P\!\!\!\!/\;\!}
\newcommand{\hqu}{\hbar}
\newcommand{\ovr}{\over}
\newcommand{\til}{\tilde}
\newcommand{\pri}{^\prime}
\renewcommand{\dag}{^\dagger}
\newcommand{\<}{\langle}
\renewcommand{\>}{\rangle}
\newcommand{\gaf}{\gamma_5}
\newcommand{\lap}{\triangle}
\newcommand{\trc}{\mathrm{tr}}
\newcommand{\al}{\alpha}
\newcommand{\be}{\beta}
\newcommand{\ga}{\gamma}
\newcommand{\de}{\delta}
\newcommand{\ep}{\epsilon}
\newcommand{\ve}{\varepsilon}
\newcommand{\ze}{\zeta}
\newcommand{\et}{\eta}
\renewcommand{\th}{\theta}
\newcommand{\vt}{\vartheta}
\newcommand{\io}{\iota}
\newcommand{\ka}{\kappa}
\newcommand{\la}{\lambda}
\newcommand{\rh}{\rho}
\newcommand{\vr}{\varrho}
\newcommand{\si}{\sigma}
\newcommand{\ta}{\tau}
\newcommand{\ph}{\phi}
\newcommand{\vp}{\varphi}
\newcommand{\ch}{\chi}
\newcommand{\ps}{\psi}
\newcommand{\om}{\omega}
\newcommand{\psb}{\overline{\psi}}
\newcommand{\etb}{\overline{\eta}}
\newcommand{\psd}{\psi^{\dagger}}
\newcommand{\etd}{\eta^{\dagger}}
\newcommand{\beq}{\begin{equation}}
\newcommand{\eeq}{\end{equation}}
\newcommand{\bdm}{\begin{displaymath}}
\newcommand{\edm}{\end{displaymath}}
\newcommand{\bea}{\begin{eqnarray}}
\newcommand{\eea}{\end{eqnarray}}
\newcommand{\mr}{\mathrm}
\newcommand{\mb}{\mathbf}
\newcommand{\Nf}{N_{\!f}}
\newcommand{\Nc}{N_{\!c}}
\renewcommand{\i}{\mathrm{i}}


\section{Introduction}

One of the main practical advantages one exploits in Monte Carlo studies of
lattice QCD is the fact that phenomenologically relevant low-energy observables
(i.e.\ masses and matrix elements of operators between physical states) may be
studied {\em without\/} a need to choose a specific gauge; i.e.\ they may be
determined from gauge-invariant correlators, and hence the functional integral
may perform a random sampling over all gauge orbits.

Nonetheless, there are good reasons to address the question how one may
transform a given ensemble of gauge configurations to a specific gauge; mainly
because on such a gauge-fixed ensemble {\em gauge-dependent\/} objects like
quark- or gluon-propagators may be determined too, and the results may be
compared with what one gets in perturbation theory or in other non-perturbative
frameworks, like the Dyson-Schwinger approach.

An attempt to fix a given configuration to a specific gauge raises several
issues.
The first question to ask is what a lattice analogue of a continuum gauge like
the Lorentz condition
\beq
\pa_\mu A_\mu = 0
\;,
\label{CLG}
\eeq
might be, and the typical answer is to construct a functional on the space of
lattice configurations such that its extremum simplifies in the continuum limit
to (\ref{CLG}), but the choice is, of course, not unique.
The simplest option is the standard lattice Landau gauge functional
\beq
F[\{U\}]={1\ovr8V}\sum_{x,\mu} \Big(
1+{1\ovr\Nc}\mr{Re}\mr{Tr}(U_\mu^\Omega(x)) \Big)
\label{LLG}
\eeq
where $U_\mu^\Omega(x\!+\!\mu)=\Omega(x)\dag\,U_\mu(x)\,\Omega_\mu(x)$ is the
gauge transformed link variable and $V\!=\!\sum_x\!1$, but it may be worthwhile
to maximize an ``improved'' functional instead (see e.g.\ \cite{Bonnet:2000bw}).
The second (more technical) question is which procedure shall be used to search
for the maximum\footnote{If there is a unique extremum at all, i.e.\ if there
are no genuine Gribov copies \cite{Gribov:1978wm}.} of a given functional like
(\ref{LLG}), since the latter represents a very high dimensional extremization
problem, and there is, unfortunately, no known algorithmic solution.

In view of this problem, Laplacian Landau gauge fixing has been proposed
\cite{LaplacianBasic} -- a method which does not directly attempt to extremize
(\ref{LLG}) but rewrites the problem such that, after a set of local
constraints is traded for a global one, the latter task may be solved exactly.
The advantage is that the procedure is {\em unique\/}, i.e.\ the final
configuration does not depend on the initial gauge.
It has led to beautiful results in quenched $SU(2)$, $SU(3)$ and $SU(4)$
studies \cite{LaplacianApplic}, but since there is no firm analytic prediction
to compare with, the situation is not conclusive.
As a result, a good fraction of the research which tries to link the QCD
confinement-deconfinement phenomenon to the behaviour of specific one- or
two-dimensional defect structures (e.g.\ monopoles or vortices) deals with
algorithmic issues and there is, to date, no agreement on the physics lesson
learned.

In this situation, we decided to address the question in a simpler setting,
namely the pure compact $U(1)$ gauge theory in 4 dimensions with the standard
Wilson action
\beq
S[\{U\}]=\be\sum_{x,\mu<\nu} \Big( 1\!-\!\mr{Re}
(U_\mu(x) U_\nu(x\!+\!\hat\mu) U_\mu(x\!+\!\hat\nu)\dag U_\nu(x)\dag) \Big)
=\be\,\sum_{\Box} \Big(1\!-\!\cos(\th_\Box)\Big)
\label{WilsonAction}
\eeq
where $\be\!=\!1/g^2$, $U_\mu(x)\!=\!\exp(\i\th_{x,\mu})\in U(1)$ and
$\th_\Box\!=\!\th_{x,\mu\nu}\!=\!\th_{x,\mu}\!+
\!\th_{x+\hat\mu,\nu}\!-\!\th_{x+\hat\nu,\mu}\!-\!\th_{x,\nu}$ is the
plaquette angle.
This theory is known to have two phases \cite{U1phases}:
For $\be\!>\!\be_\mr{c}$ it is in the ordered (``Coulomb'') phase, for
$\be\!<\!\be_\mr{c}$ it is in the disordered (``confined'') phase.
Contrary to what early and some more recent studies indicate
\cite{U1secondorder}, the transition seems to be 1st order \cite{U1firstorder}
with $\be_\mr{c}\!\simeq\!1.01$.

What is important in our context is that the $U(1)$ theory, if quantized on the
torus $T^4$, shows the Gribov ambiguity mentioned above
\cite{Killingback:1984en}.
Several numerical studies have tried to pin down the effect gauge fixing
artefacts have on the mass of the ``photon'' in either phase.
First, Coddington, Hey, Mandula and Ogilvie studied the transverse
zero-momentum correlator
\beq
C(t)=\sum_{\mb{x},\mb{y}}\sum_{i=1}^3 A_i(\mb{x},0) A_i(\mb{y},t)
\eeq
where they defined the photon field (and we follow them in this choice) through
\beq
A_\mu(x)\!=\!\mr{Im}\,U_\mu(x)\!=\!\sin(\th_{x,\mu})
\;,
\label{photdef}
\eeq
and they found that --~at zero momentum~-- the ``photon'' behaves as a massless
particle for $\be\!>\!\be_\mr{c}$ but as a massive one at $\be\!<\!\be_\mr{c}$
\cite{Coddington:1987yz}.
Thereby they confirmed an earlier result, gotten in a gauge-invariant approach
\cite{BergCox}.
Nakamura and Plewnia extended this work to arbitrary spatial momentum,
i.e.\ they measured the correlator 
\beq
C_{\mu\nu}(\mb{p},t)=\sum_{\mb{x},\mb{y}}A_\mu(\mb{x},0) A_\nu(\mb{y},t)
\,e^{-\i\mb{p}\cdot(\mb{x\!-\!y})}
\label{cmom}
\eeq
for a few low-lying $\mb{p}$ and observed that the transverse part at
$\be\!=\!1.1$ deviates, for large time separations, quite drastically from the 
perturbative prediction, if (\ref{LLG}) is maximized through local iterative
steps only \cite{Nakamura:1991ww}.
They also noticed that the result improves, if the local iterative procedure is
extended to opt for the best one out of several attempts, and this observation
triggered a long stream of subsequent activities (see below).

The goal of the present study is to give a fair comparison between Laplacian
and iterative gauge-fixing in the pure $U(1)$ theory, where both phases may be
considered, good statistics is easy to obtain and correlators involving all
momenta may be determined through fast Fourier transform.
After a quick review of the gauge-fixing procedures, the comparison is given
for the Coulomb and confined phases separately, and we end with a short
discussion of the relevance of our results to the non-Abelian case.


\section{Iterative versus Laplacian gauge fixing}

We start with a brief outline of the two gauge fixing algorithms which we are
going to compare w.r.t.\ the photon correlators implied.

\subsection{Standard and modified iterative Landau gauge fixing}

Modified iterative Landau gauge fixing is an improvement on standard iterative
Landau gauge fixing.
The latter is tantamount to iteratively passing through the lattice and trying
to {\em locally maximize\/} the gauge functional (\ref{LLG}) which in the case
of the compact $U(1)$ theory reads
\beq
F[\{\th\}]={1\ovr V}\sum_x F_x(\th)
\;,\quad
F_x(\th)={1\ovr16} \sum_{\mu=1}^4
\Big( 2+\cos(\th_{x,\mu})+\cos(\th_{x-\hat\mu,\mu}) \Big)
\;.
\label{func}
\eeq
In this symmetrized version it is obvious that for a given site $x$ the local
gauge transformation
\beq
\Omega(x):\th_{x,\mu}\to\th_{x,\mu}\!+c_x\;,\;\;
\th_{x-\hat\mu,\mu}\to\th_{x-\hat\mu,\mu}\!-c_x
\label{loctrf}
\eeq
which maximizes $F_x(\th)$ may be found analytically.
Moreover, doubling the respective $c$ leaves the local contribution $F_x(\th)$
unchanged, thus offering the possibility of an overrelaxation step: in practice
one sweeps through the lattice in a mixed overrelaxation/maximization strategy
\cite{deForcrand:1989im}.

In general, this standard procedure gets stuck in a local extremum.
If the common belief that the ``true'' gauge copy corresponds to the global
maximum \cite{Zwanziger} is correct, an extra effort is needed to improve on
this situation.
Specifically for the case of the compact $U(1)$ theory two types of objects are
known which prevent the local iterative procedure from reaching the true
extremum.
Bogolubsky et al.\ have shown \cite{Bogolubsky:1999cb} that in the quest for
the global maximum one needs to suppress both the ``double Dirac sheets'' (DDS)
and the ``zero momentum modes'' (ZMM) of the gauge field, two concepts which we
shall explain in detail.

DDS:
Each plaquette angle $\th_{x,\mu\nu}$ is decomposed into the gauge-invariant
electromagnetic flux $\overline\th_{x,\mu\nu}\in\,]\!-\!\pi,\pi[$ and the
discrete gauge-dependent part $2\pi n_{x,\mu\nu}, n_{x,\mu\nu}=0,\pm1,\pm2$
\cite{DeGrand:eq}.
A value $n_{x,\mu\nu}=\pm1$ represents a Dirac string passing through the
plaquette and a plaquette with $n_{x,\mu\nu}\neq0$ is called a ``Dirac
plaquette''.
A gauge-invariant monopole inside an elementary cube is detected via its
non-zero flux $2\pi \sum_6 n_{\mu\nu}$. Since monopole worldlines form closed
loops on the dual lattice, the Dirac strings normally span cylindrical $2d$
sheets dual to the Dirac plaquettes.
However, on a lattice with periodic boundary conditions, it is also possible
to form a boundary-free Dirac sheet spanning a whole 2-torus. It represents
the exceptional case of two monopole loops merging on top of each other.
A pair of Dirac sheets with opposite flux orientation carries zero action,
and can be removed by a periodic gauge-transformation \cite{Grosch:1985cz}.
It is called a ``double Dirac sheet'' (DDS).
In the confining phase, monopole worldlines percolate \cite{percolation}, and
so do Dirac strings and sheets. In the Coulomb phase, monopole loops are small,
Dirac plaquettes are less abundant and DDS are rare. When they appear,
they tend to lie parallel to the main axes, presumably for entropic reasons,
since this orientation requires fewer Dirac plaquettes.
In any case, a DDS in the $\si\rh$ plane may be detected by counting
the total number of $\mu\nu$ Dirac plaquettes
($\varepsilon_{\mu\nu\si\rh} \neq 0$), which must satisfy
\beq
N_\mr{DP}^{(\mu\nu)} \geq 2L_\si L_\rh
\label{DDScond1}
\;.
\eeq

ZMM:
With periodic boundary conditions the ``zero momentum modes'' of the gauge
field
\beq
\overline\th_\mu={1\ovr V}\sum_x \th_{x,\mu}
\label{ZMM}
\eeq
do not contribute to the action (\ref{WilsonAction}) -- the angle of the
Polyakov loop is arbitrary, indicating a global $U(1)$ symmetry of the action.
For gauge configurations representing small fluctuations around constant modes
it is easy to see that the global extremum of the functional (\ref{func})
requires $\overline\th_\mu=0$.
The latter condition may be satisfied through a {\em non-periodic\/}
transformation
\beq
\th_{x,\mu}\to\th_{x,\mu}^c\equiv\th_{x,\mu}-c_\mu\;\;\mr{mod}\;2\pi
\;\;,\;\;c_\mu\in\,]\!-\!\pi,\pi[
\label{NonperiodProposal}
\eeq
with the specific choice $c_\mu=\overline\th_\mu$, but in general this is not a
gauge-transformation (the Polyakov loop is changed).
If one insists that (\ref{NonperiodProposal}) shall be a gauge-transformation,
the $c_\mu$ need to be chosen from the (finite) set
\beq
c_\mu\in\{0,\pm{2\pi\ovr L_\mu},\pm{4\pi\ovr L_\mu},\ldots\}
\;.
\label{NonperiodRestriction}
\eeq
Had we chosen the other natural definition of the gauge field,
$A_\mu(x)\!=\!-\i\log(U_\mu(x))\!=\!\th_{x,\mu}$ rather than (\ref{photdef}),
the transformation (\ref{NonperiodProposal}) with $c_\mu=\overline\th_\mu$
would amount to a shift in $A_\mu(x)$ such that its constant Fourier mode is
zero.
In other words, abandoning the restriction (\ref{NonperiodRestriction}) amounts
to eliminating a constant background field.
While this is frequently done (see e.g.\ \cite{Bogolubsky:1999cb}),
we feel that sticking with the combination (\ref{NonperiodProposal},
\ref{NonperiodRestriction}), where the background is suppressed but not
eliminated, is somewhat closer in spirit to the original extremization task --
choosing, out of (\ref{NonperiodRestriction}), a non-trivial $c_\mu$, amounts
to switching to another point on the gauge orbit which, in view of the global
extremum, looks more promising.
In practice, one does not expect any sizable difference between the two
options, since on a large enough lattice the discreteness of $c_\mu$ turns into
a rather marginal restriction.

The ``modified iterative Landau gauge'' (modified ILG) differs from the
``standard'' ILG discussed above in that it tries to eliminate both the DDS and
the ZMM as far as possible.
To implement it, we proceed in bunches of 100 standard sweeps through the
lattice followed by a ZMM operation (\ref{NonperiodProposal},
\ref{NonperiodRestriction}) and a check on the DDS condition (\ref{DDScond1})
which we sharpen to
\beq
N_\mr{DP}^{(\mu\nu)}\geq 1.5L_\si L_\rh
\label{DDScond2}
\eeq
to account for additional monopole-antimonopole pairs which may punch holes
into the sheet.
In case (\ref{DDScond2}) is met, the whole gauge-fixing procedure is started
over from another random point on the gauge orbit.
This is repeated until a configuration without DDS is obtained.
Further bunches of 100 extremization/overrelaxation sweeps are applied until the
last round has increased the functional (\ref{func}) by less than $10^{-8}$.
To prevent the program from getting trapped in an infinite loop, a cut-off on
the outer loop is introduced which is chosen $O(30)$ in the Coulomb phase and
$O(10)$ in the confined phase (these figures are tuned to our $\be$ values,
i.e.\ they represent an ad-hoc element in the procedure).
In such a case the result of the random gauge copy which has led to the
highest functional value (\ref{func}) is post-iterated with local
maximization/overrelaxation steps to make sure that at least a local extremum
is reached.

For reasons which will become clear later on, we decided to include the ZMM
suppression (\ref{NonperiodProposal}, \ref{NonperiodRestriction}) in our
``standard'' ILG ensemble, too.

\subsection{Laplacian Landau gauge fixing}

The Laplacian Landau gauge fixing method \cite{LaplacianBasic} takes its origin
from the observation that maximizing (\ref{func}) (we specify the discussion to
the $U(1)$ case) is equivalent to {\em minimizing\/}
\beq
G[\{\th\}]=\mr{Re}\big(
\sum_{x,\mu}
2
-e^{\i\th_{x,\mu}}
-e^{-\i\th_{x-\hat\mu,\mu}}
\big)
\label{minigeneral}
\eeq
and the latter means, more explicitly, to search for a gauge transformation
(\ref{loctrf}) such that
\beq
G[\{\th\}]=\mr{Re}\big(
\sum_{x,\mu}
 e^{\i c_x} 2 e^{-\i c_x}
-e^{\i c_x} e^{\i\th_{x,\mu}} e^{-\i c_{x+\hat\mu}}
-e^{\i c_x} e^{-\i\th_{x-\hat\mu,\mu}} e^{-\i c_{x-\hat\mu}}
\big)
\label{miniexplicit}
\eeq
is minimal. Setting $\ph_x\!=\!\rh_x e^{-\i c_x}$, the bracket suggestively
looks like the quadratic form
\beq
\sum_{x,y} \ph_x\dag(-\triangle)_{xy}\ph_y
\label{covlap}
\eeq
and the problem (\ref{miniexplicit}) would, in fact, be equivalent to searching
for the {\em lowest eigenmode\/} of the sign flipped covariant Laplacian, were
it not for the set of local constraints $|\ph_x|\!=\!1$ (or $\rh_x\!=\!1$)
for all $x$.
The Laplacian Landau gauge \cite{LaplacianBasic} ignores the latter and
constructs the transformation (\ref{loctrf}) from the eigenmode which
corresponds to the lowest eigenvalue of (\ref{covlap}), i.e.\ it sets
$c_x\!=\!-\mr{Im}\log(\ph_x)$ and keeps only the global constraint
$\sum_x |\ph_x|^2\!=\!1$ (because the mode may be normalized).

The Laplacian Landau gauge (LLG) is unique, i.e.\ the final result does not
depend on the initial gauge \cite{LaplacianBasic}, and it is unambiguous in
all cases where (a) the lowest eigenmode of the covariant Laplacian is
non-degenerate and (b) the result satisfies $\ph_x\!\neq\!0$ for all sites $x$.
For a continuous (i.e.\ interpolating) gauge field $\til A_\mu(\til x)$, the
local ambiguities $\til\ph(\til x)\!=\!0$ are associated with Dirac sheets.
For details with respect to (a/b) we refer to \cite{LaplacianBasic,
LaplacianApplic} and for a discussion of the relationship of LLG to Landau
gauge in perturbation theory we refer to \cite{vanBaalMandula}.
To compute the lowest eigenmode of the covariant Laplacian we use the Arnoldi
package ARPACK \cite{arpack}.


\section{Comparison in the Coulomb phase}

We generate 1500 configurations on a $16^4$ lattice at $\beta\!=\!1.05$, fix
them with the Laplacian and the modified iterative prescription and compare the
results for the photon propagator.

\begin{figure}[t]
\vspace{-2mm}
\begin{center}
\epsfig{file=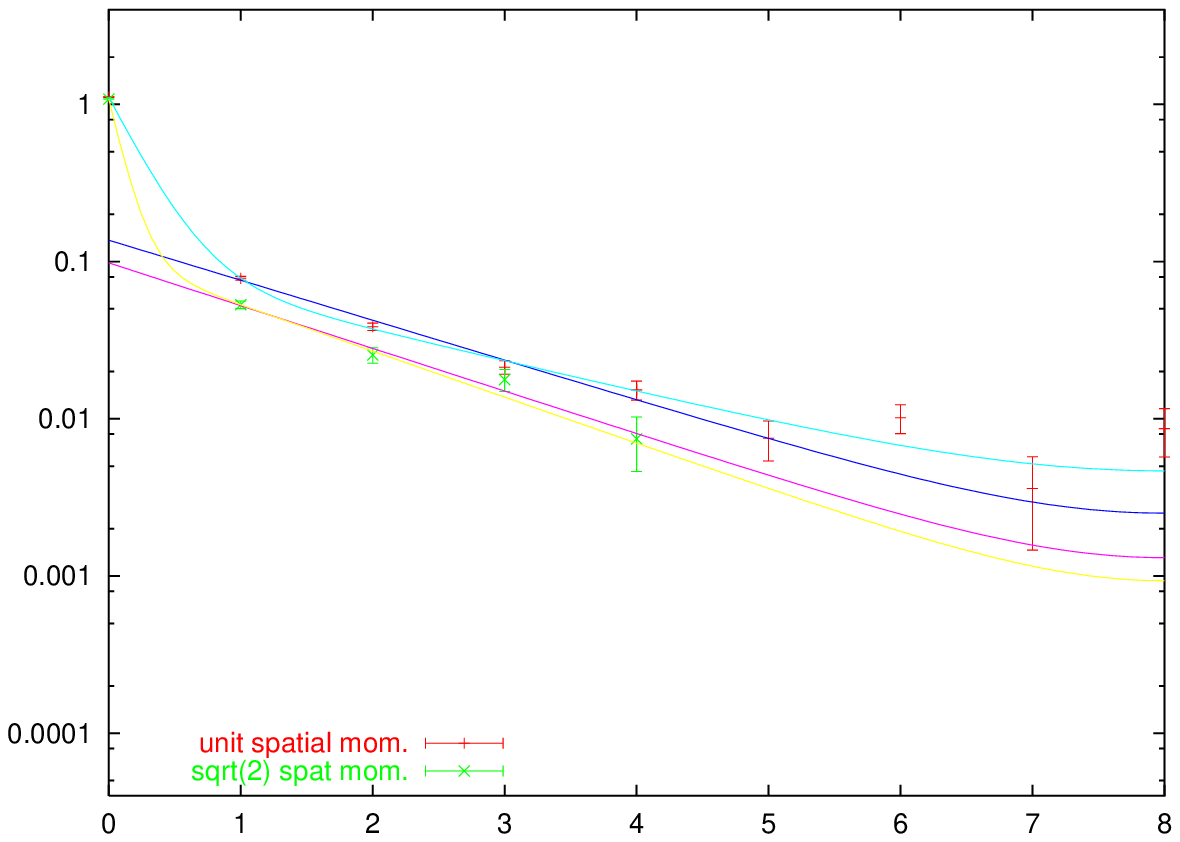,height=6cm,width=8.6cm,angle=0}
\\
\epsfig{file=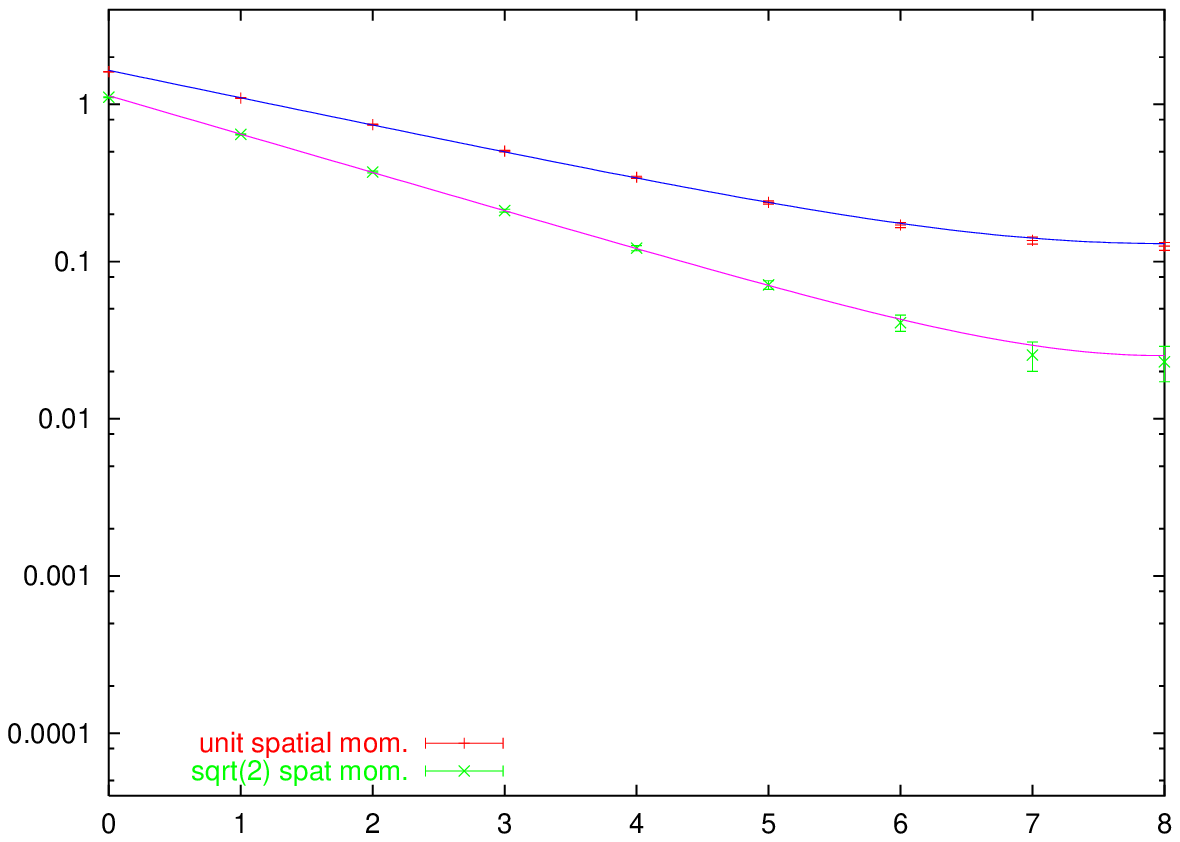,height=6cm,width=8.6cm,angle=0}
\hfill\hspace{-5mm}\hfill
\epsfig{file=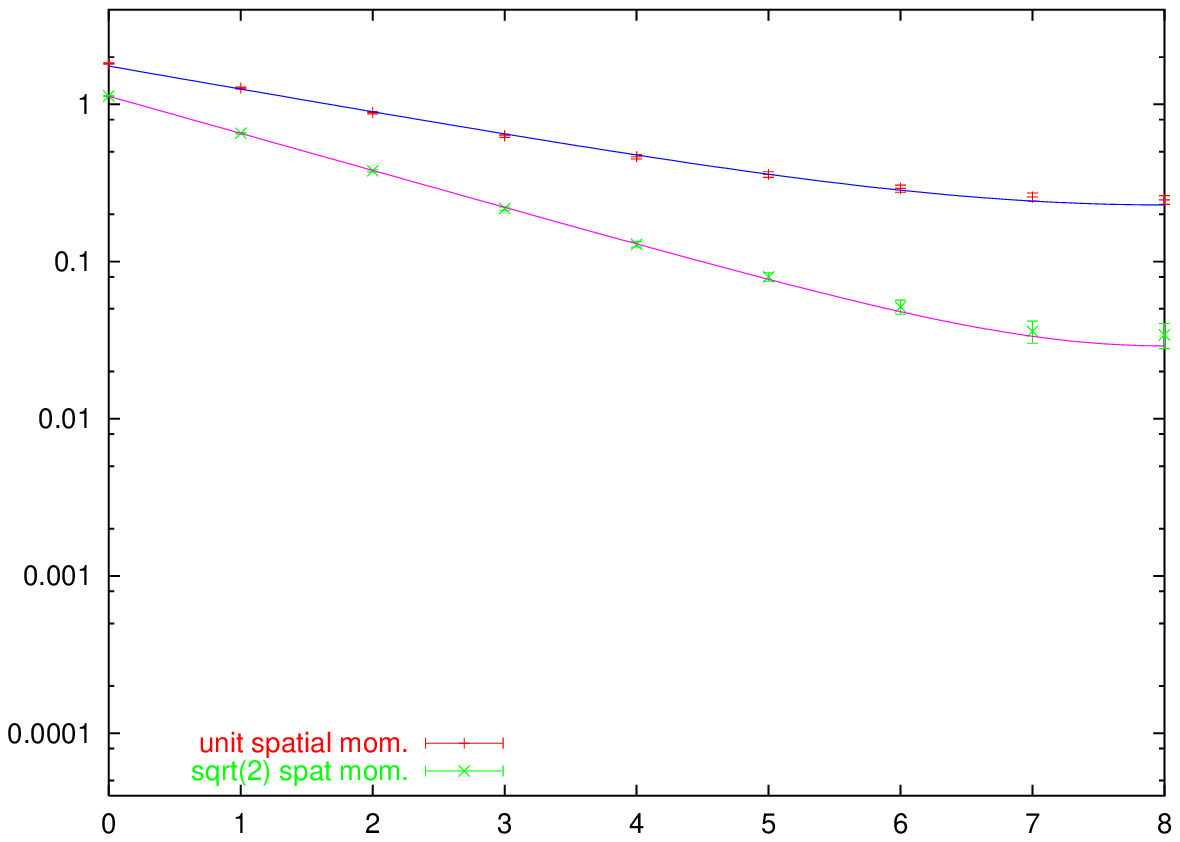,height=6cm,width=8.6cm,angle=0}
\end{center}
\vspace{-8mm}
\caption{\sl\small
Top: Correlator between planes of Polyakov loops with once $(+)$ or $\sqrt{2}$
times the minimum spatial momentum at $\be\!=\!1.05$. From these one might
determine two points of the photon dispersion relation $(\ref{meff})$, but this
gauge-invariant estimator is rather noisy.
The same information may be gotten, after gauge-fixing, with considerably
reduced error bars from the photon-photon correlator with the respective
momenta in the modified iterative $($bottom left$)$ or Laplacian Landau gauge
$($bottom right$)$.}
\end{figure}

A gauge-invariant way to establish, for $\be\!>\be_\mr{c}$, that the photon is
massless is to measure the (exponential) decay of the correlator between two
planes of spatial Polyakov loops with a few nonzero momenta in the spatial
directions.
At large enough Euclidean time separation $t$ one determines the effective
mass $m_\mr{eff}$.
After introducing the lattice momentum
\beq
\hat p_\mu={2\ovr a}\sin({p_\mu a\ovr2})={2\ovr a}\sin({\pi n_\mu\ovr L_\mu})
\label{plat}
\eeq
and plotting $m_\mb{eff}$ as a function of $\hat\mb{p}^2$ and fitting with
\beq
m_\mr{eff}\!=\!\sqrt{m^2\!+\!\hat\mb{p}^2}
\,,
\label{meff}
\eeq
the data ought to be consistent with $m^2\!=\!0$.
Since the available (ordinary) momenta are $\mb{p}_i\!=\!2\pi\,\mb{n}_i/L_i$
and good agreement with the dispersion relation (\ref{meff}) typically requires
$\mb{p}^2\!\ll\!1$, the spatial extension of the lattice needs to be large.
On our $16^4$ lattice, results are acceptable for the two lowest momenta,
$|\mb{p}|\!=\!1\!\cdot\!2\pi/16$ and $|\mb{p}|\!=\!\sqrt{2}\!\cdot\!2\pi/16$,
as displayed in Fig.\ 1 (top).
We fit the correlator with a single and a double $\cosh$ and check whether the
two effective masses are consistent with (\ref{meff}) where the mass has been
set to zero, which indeed they are.

\begin{figure}[t]
\vspace{-2mm}
\epsfig{file=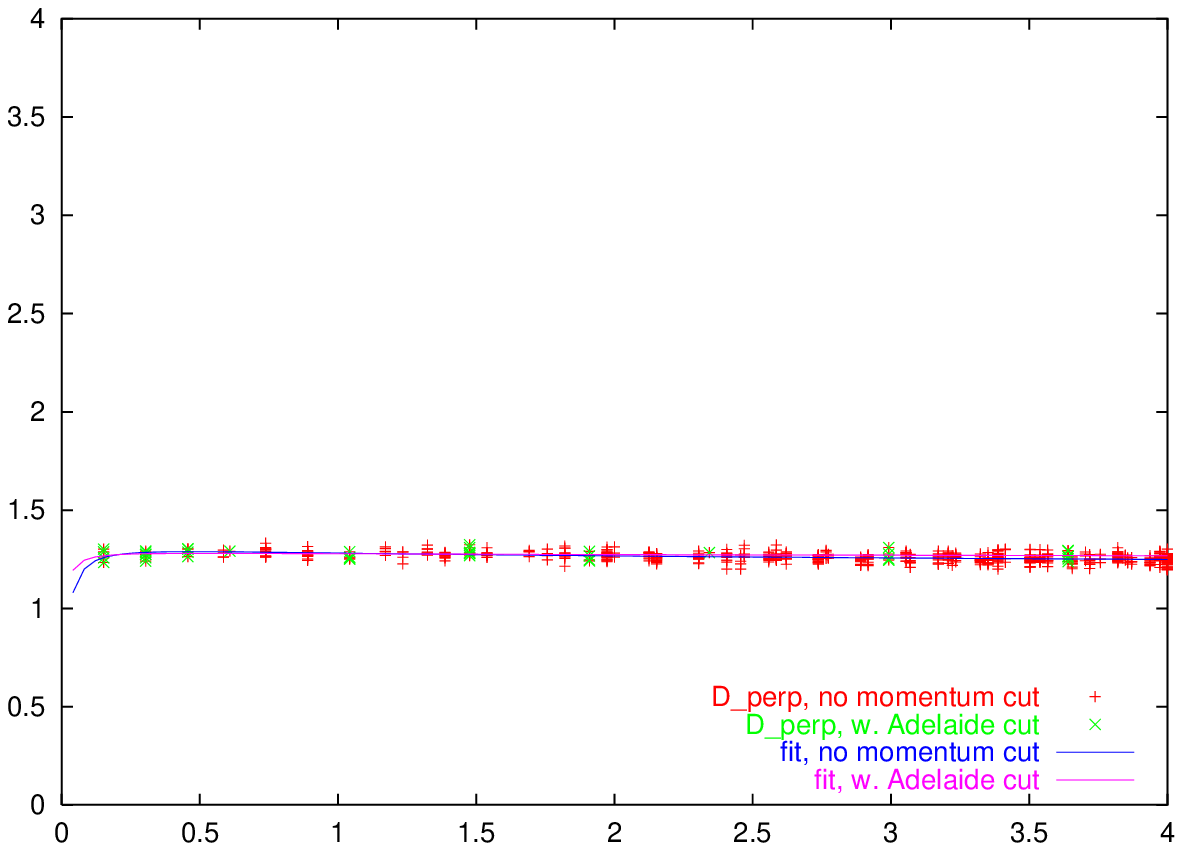,height=6cm,width=8.6cm,angle=0}
\hfill\hspace{-5mm}\hfill
\epsfig{file=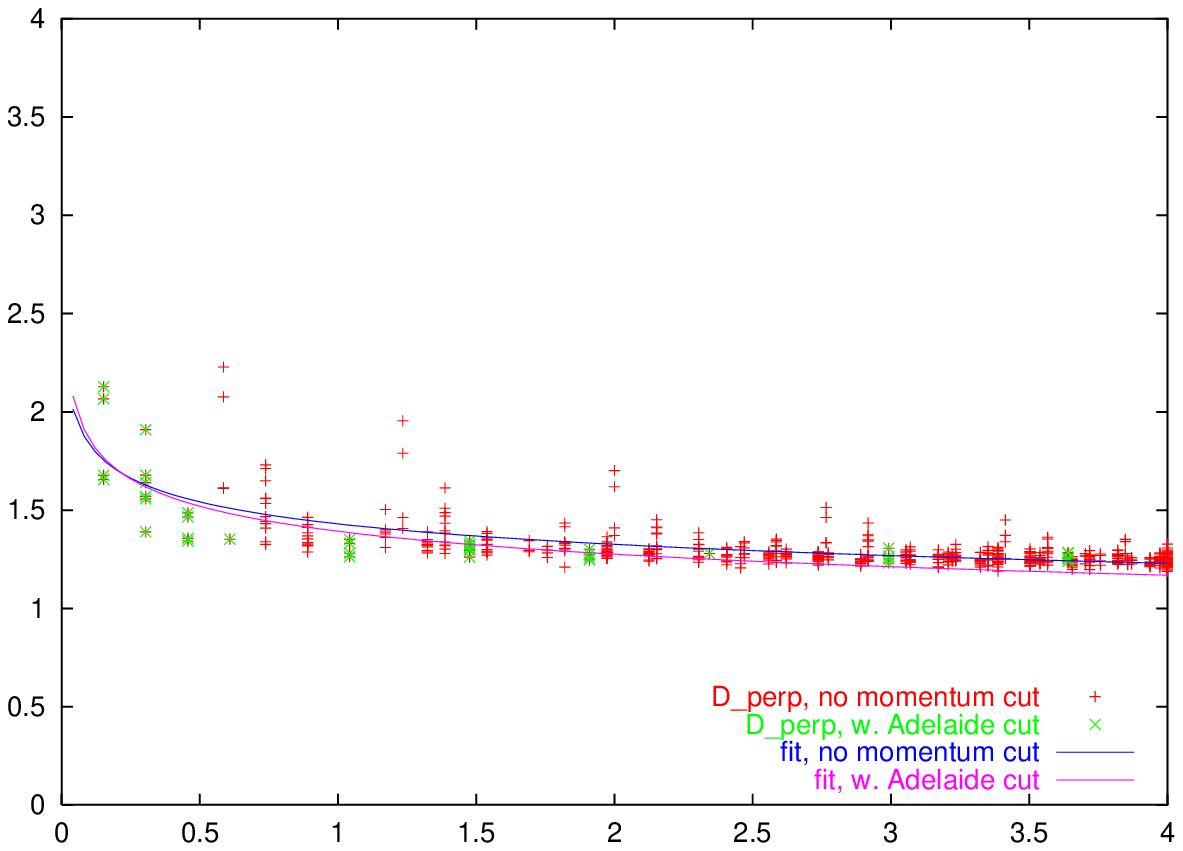,height=6cm,width=8.6cm,angle=0}
\\
\epsfig{file=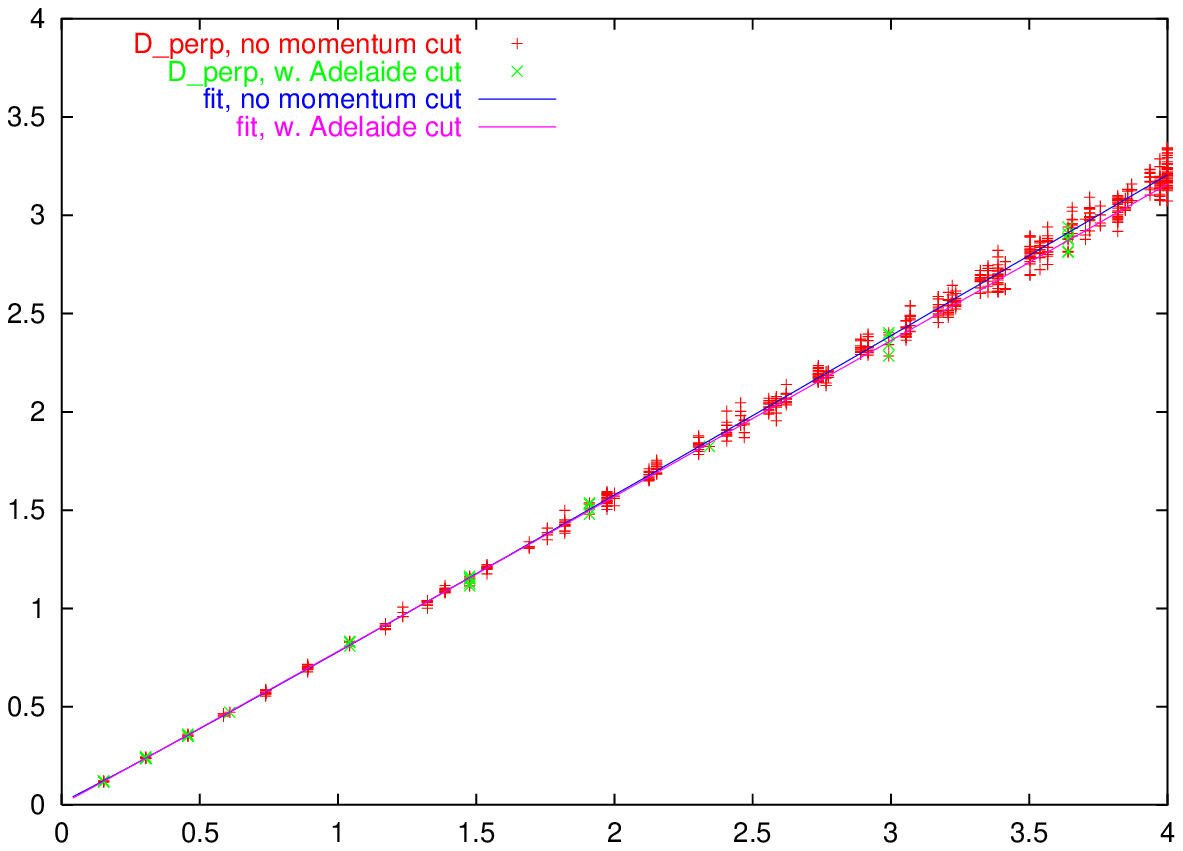,height=6cm,width=8.6cm,angle=0}
\hfill\hspace{-5mm}\hfill
\epsfig{file=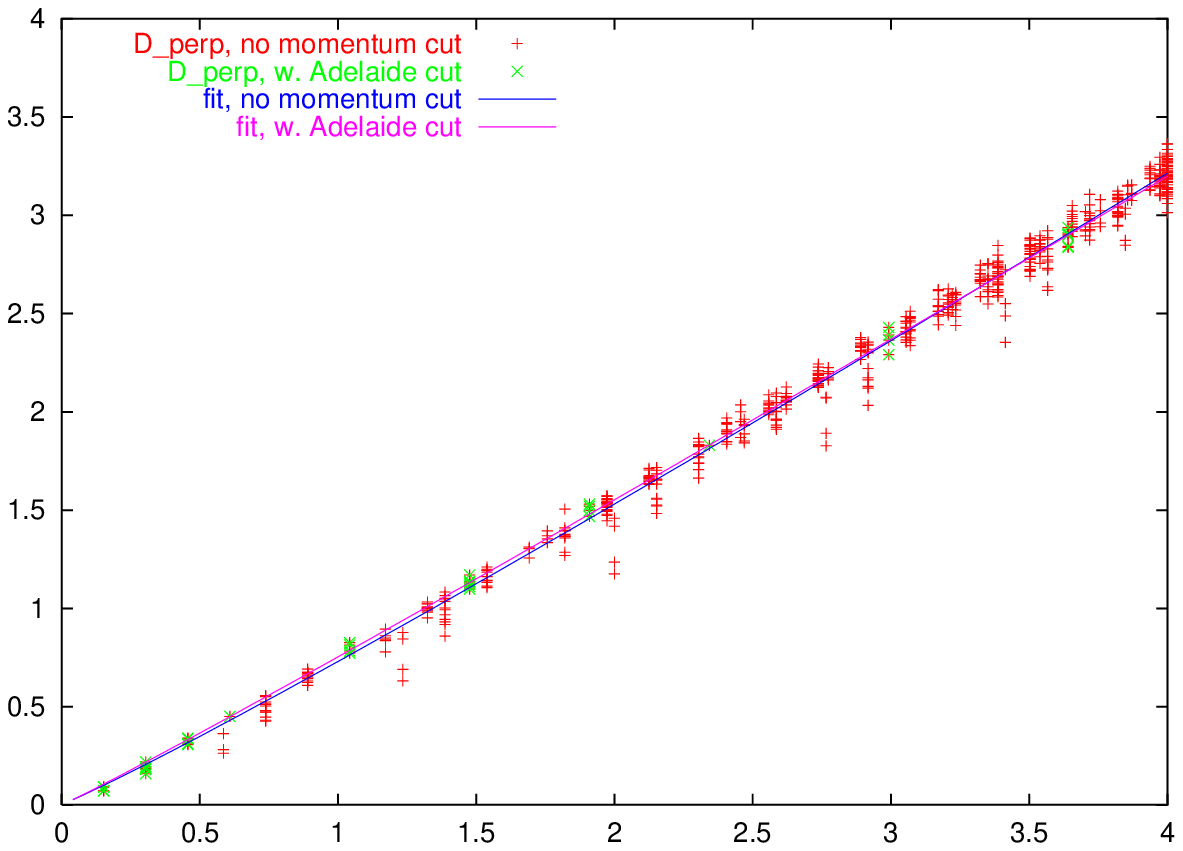,height=6cm,width=8.6cm,angle=0}
\vspace{-8mm}
\caption{\sl\small
Transverse component of the photon propagator at $\beta\!=\!1.05$ on a $16^4$
lattice, after modified iterative $($left$)$ and Laplacian $($right$)$ Landau
gauge fixing. Since perturbatively $D_\perp(p^2)\!\simeq\!1/p^2$, the graphs
$\hat p^2 D_\perp(\hat p^2)$ vs.\ $\hat p^2$ $($top$)$ and $1/D_\perp(\hat p^2)$
vs.\ $\hat p^2$ $($bottom$)$ are supposed to show an almost linear dependence.
The modified ILG shows good rotational invariance, while the LLG shows a huge
scatter of the data at a given $\hat p^2$ $(+)$. After the cylindrical momentum
cut $(\times)$ the scatter on the r.h.s.\ is reduced $($i.e.\ rotational
invariance improves$)$, but the IR enhancement in the Laplacian case persists.}
\end{figure}

Following Nakamura and Plewnia \cite{Nakamura:1991ww}, we repeat this in a
gauge-dependent way, using our two gauge-fixed ensembles.
To that aim we determine the decay properties of the transverse photon-photon
correlator (\ref{cmom}), again for the same spatial momenta and with indices
$\mu\!=\!\nu$ which point orthogonal to both $\mb{p}$ and the propagation
direction (associated with $t$).
The result is displayed in Fig.\ 1 (bottom).
The important point is that the link-link correlator yields the same
information about the photon dispersion relation as the Polyakov loop
correlator, but with much higher accuracy.
This is easy to understand, since the main effect of the gauge transformations
is to ``unwind'' the Polyakov loop, i.e.\ to remove multiples of $2\pi$ which
only contribute to the noise.

The main technical point of the present paper is that one may get more detailed
information about the performance of either fixing procedure by considering the
complete correlator
\beq
C_{\mu\nu}(p)=\sum_{x,y}A_\mu(x) A_\nu(y)\,e^{-\i p(x\!-\!y)}
=\hat A_\mu(p) \hat A_\nu(-p)
\label{carb}
\eeq
for arbitrary four-momentum $p$.
Since in the continuum the photon propagator is transverse, it makes sense to
split it (on the lattice) into a transverse and a longitudinal piece, viz.\
\beq
C_{\mu\nu}(\hat p)=
(g_{\mu\nu}-{\hat p_\mu \hat p_\nu\ovr \hat p^2})D_\perp+
{\hat p_\mu \hat p_\nu\ovr \hat p^2}D_\parallel
\label{splitdef}
\eeq
with
\beq
D_\perp(\hat p^2)={1\ovr3}
\Big(\trc(C_{\si\rh})-{\hat p_\si C_{\si\rh} \hat p_\rh \ovr \hat p^2}\Big)
\quad,\qquad
D_\parallel(\hat p^2)={\hat p_\si C_{\si\rh} \hat p_\rh \ovr \hat p^2}
\;.
\label{splitfun}
\eeq
If the propagator is transverse (i.e.\ $D_\parallel(\hat p^2)\!=\!0$), the
transverse coefficient $D_\perp(\hat p^2)$ is simply obtained as $1/3$ of the
trace of the propagator $C_{\mu\nu}(\hat p)$.

\begin{figure}[t]
\vspace{-2mm}
\epsfig{file=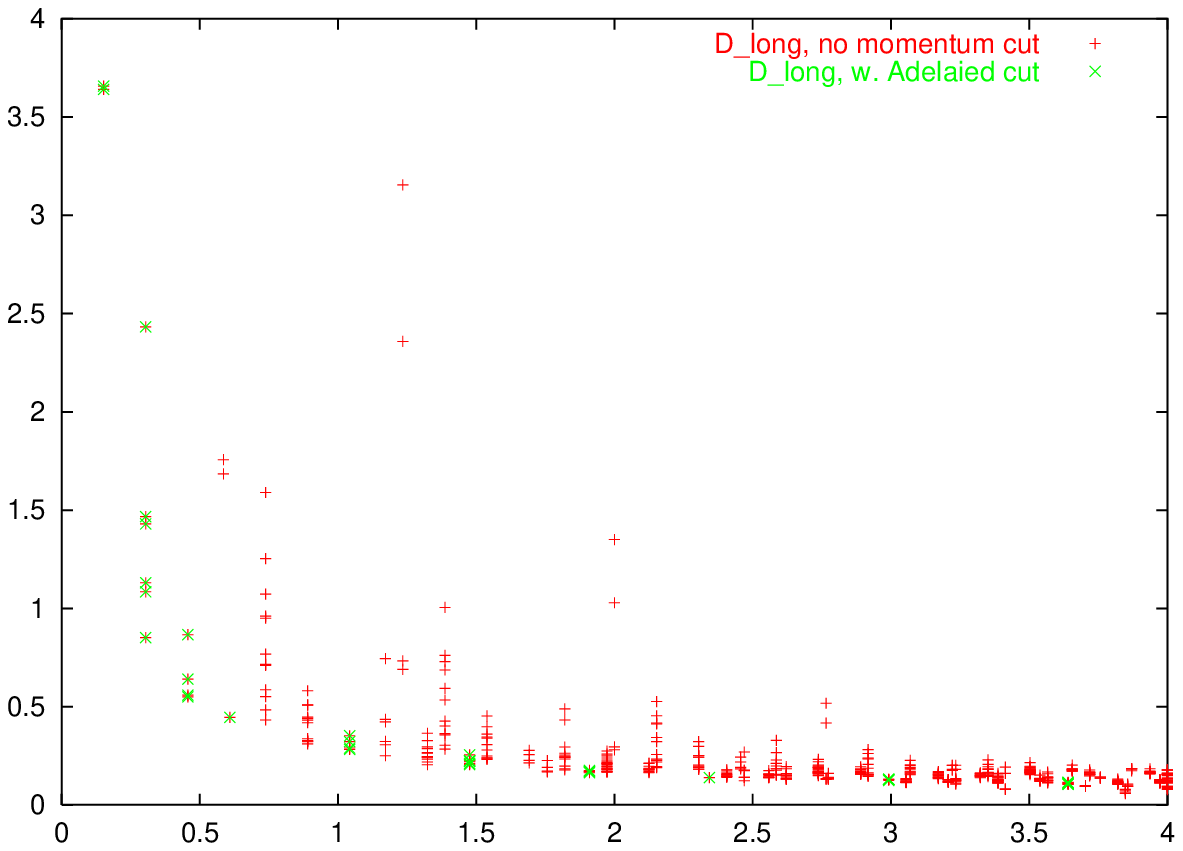,height=6cm,width=8.6cm,angle=0}
\hfill\hspace{-5mm}\hfill
\epsfig{file=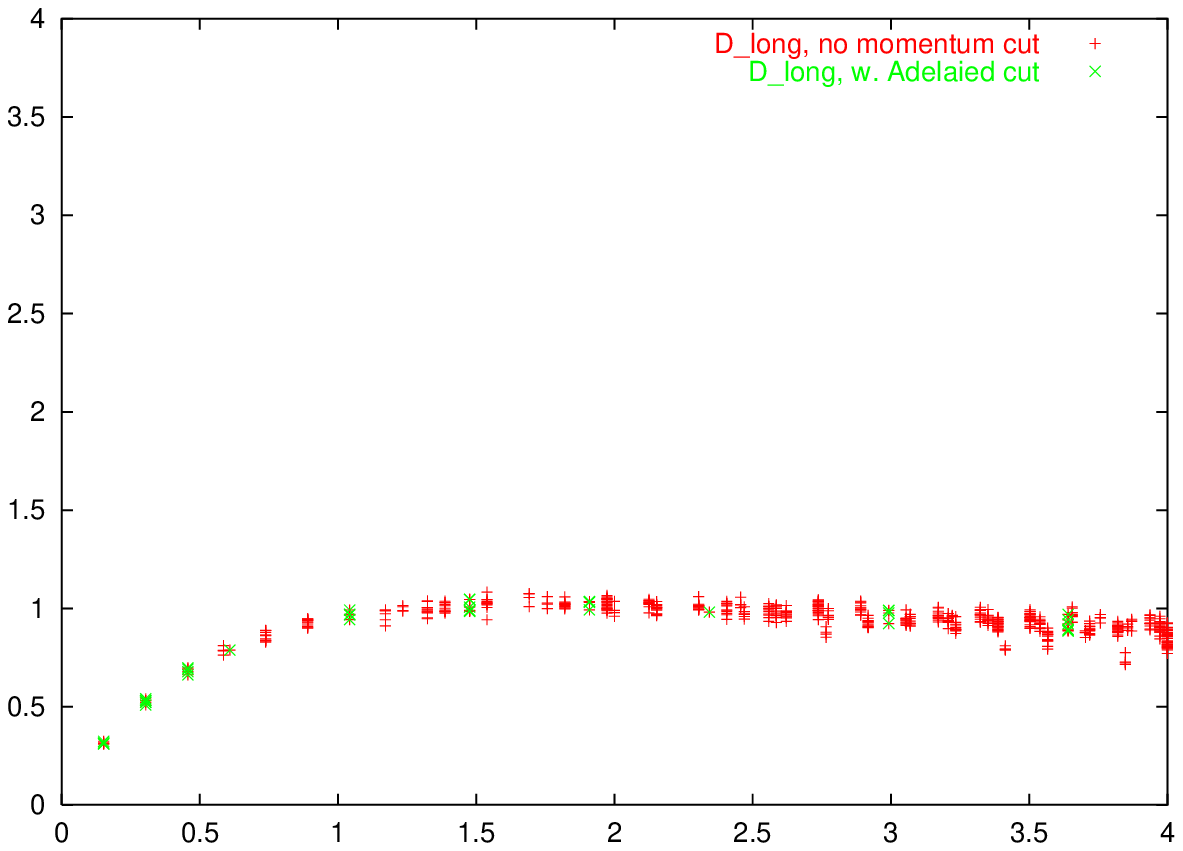,height=6cm,width=8.6cm,angle=0}
\vspace{-8mm}
\caption{\sl\small
The longitudinal part of the correlator (\ref{carb}), multiplied by $\hat p^2$,
in the Laplacian gauge at $\be\!=\!1.05$ $($l.h.s.$)$ and at $\be\!=\!0.95$
$($r.h.s.$)$. This confirms earlier findings that the LLG does not yield a
purely transverse propagator, but the longitudinal admixture seems to be much
more severe $($in particular for small momenta$)$ in the Coulomb phase
$($l.h.s.$)$ than in the confined phase $($r.h.s.$)$.}
\end{figure}

In Fig.\ 2 we display our results for the Coulomb phase.
Since in the weak coupling limit $D_\perp(p^2)\simeq1/p^2$, we plot
$\hat p^2 D_\perp(\hat p^2)$ vs.\ $\hat p^2$ (top) and $1/D_\perp(\hat p^2)$
vs.\ $\hat p^2$ (bottom) -- both of them after modified iterative (left) and
Laplacian (right) gauge-fixing.
The striking fact is the huge amount of scatter in the data for the Laplacian
case -- data for a given $\hat p^2$ do not collapse on a universal line, i.e.\
{\em rotational invariance\/} is seriously broken after Laplacian fixing in
the Coulomb phase.
A simple way to elaborate on this point is to apply a cylindrical momentum
cut \cite{Leinweber:1999uu}, i.e.\ to decompose each momentum according to
$\hat p=p_\parallel \hat p_0+\hat p_\perp$ where $\hat p_0={1\ovr2}(1,1,1,1)$
and to retain only those momenta with a component orthogonal to $(1,1,1,1)$
that is smaller than some constant, e.g.\ $|\hat p_\perp|^2<0.2$.
Those momenta out of the full set $(+)$ which survive this cut are plotted with
crosses ($\times$).
For large momenta the scatter is considerably reduced, but for the very low
$\hat p^2$ the cut gets less effective and the IR enhancement by which the
Laplacian correlator differs from its modified iterative counterpart is
unchanged.
Towards the end of this article evidence will be presented that the Laplacian
gauge suffers from DDS, and the present observation is consistent with that
claim: If DDS persist in the Laplacian ensemble, then one would expect
rotational symmetry to be broken (remember that DDS prefer to be planar) and
expect low-lying momenta to be particularly affected, since DDS represent
large-scale objects.

In order to make contact with other studies, we fit the correlator to a
functional form which allows for some IR enhancement; we use
\beq
D(q^2)=Z\;{m^{2\al}\ovr q^{2(1+\al)}+m^{2(1+\al)}}
\label{propansatz}
\eeq
with a ``photon mass'' $m$ and an ``anomalous dimension'' $\al$ (apart from the
renormalization factor $Z$).
Then our results may be stated as follows: In the Coulomb phase the modified
iterative prescription is consistent with a propagator (\ref{propansatz}) with
zero mass and vanishing anomalous dimension, whereas the Laplacian gauge
suggests a {\em positive anomalous dimension\/}.

An interesting point is to check the transversality of the propagator -- we
find it well obeyed in the modified ILG but seriously violated in the LLG.
Fig.\ 3 shows $\hat p^2 D_\parallel$ vs.\ $\hat p^2$ for the Laplacian case
both in the Coulomb phase (l.h.s.) and in the confined phase discussed below.


To test whether the modified iterative and the Laplacian prescriptions differ
on a deeper level (i.e.\ before the photon propagator is computed), we checked
the distribution of link angles.
They produce almost indistinguishible distributions (which are very close to a
Gaussian centered about zero), i.e.\ they both comply with the perturbative
preference for zero link angle.


\section{Comparison in the confined phase}

We generate 1500 configurations on a $16^4$ lattice at $\beta\!=\!0.95$, fix
them with the Laplacian and the modified iterative prescription and compare the
results for the correlator (\ref{carb}, \ref{photdef}).

\begin{figure}[t]
\vspace{-2mm}
\epsfig{file=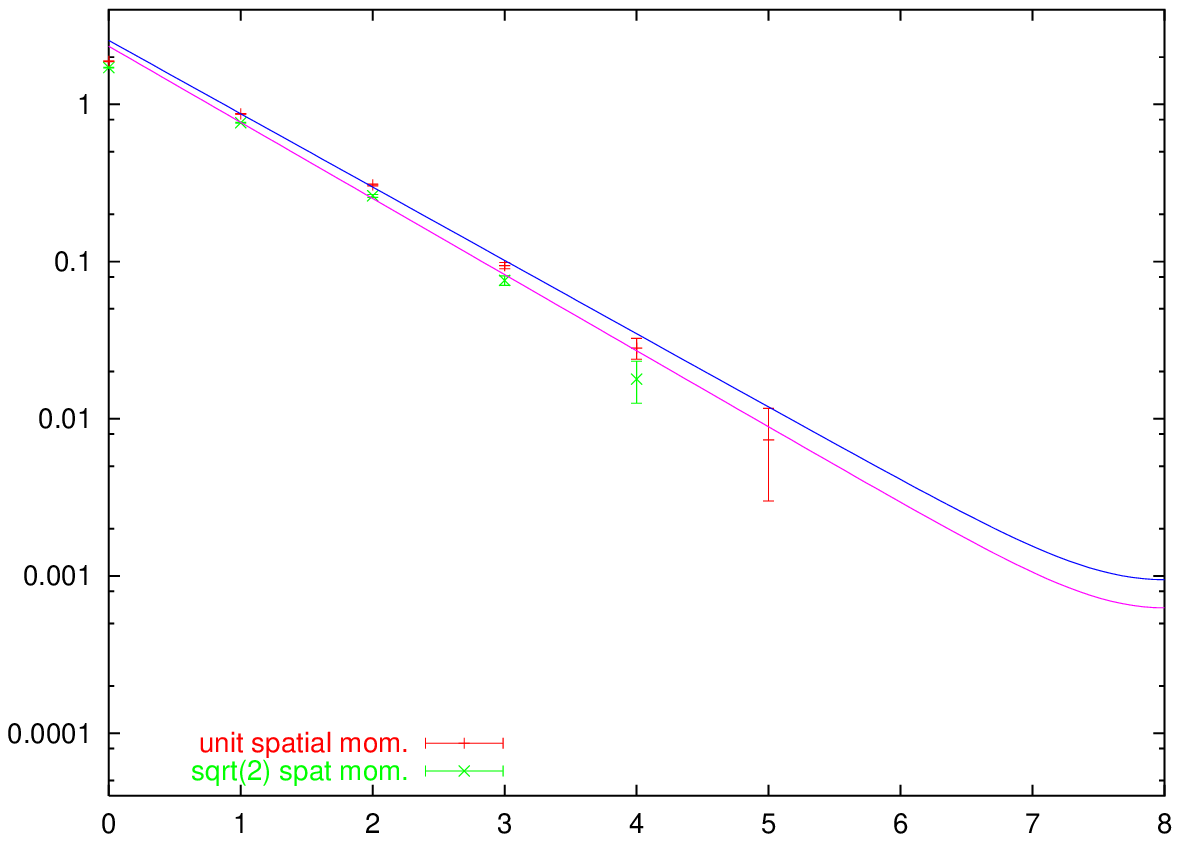,height=6cm,width=8.6cm,angle=0}
\hfill\hspace{-5mm}\hfill
\epsfig{file=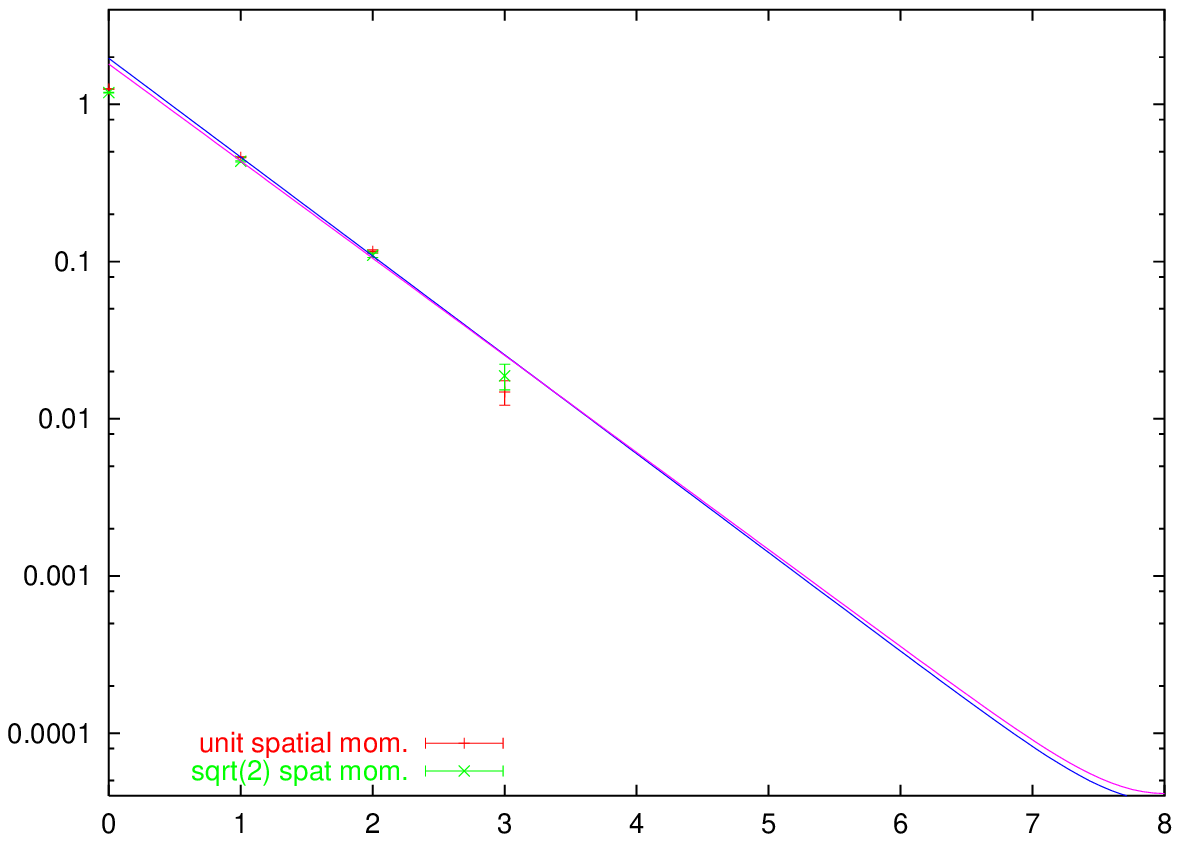,height=6cm,width=8.6cm,angle=0}
\vspace{-8mm}
\caption{\sl\small
Field (\ref{photdef}) correlator with with once $(+)$ or $\sqrt{2}$ times the
minimum spatial momentum at $\be\!=\!0.95$. This yields information about two
points in the dispersion relation $(\ref{meff})$. In the confined phase the
Polyakov loop correlator is so noisy that no signal can be extracted. The
effective mass in the modified iterative gauge $($l.h.s.$)$ is smaller than
that in the Laplacian gauge $($r.h.s.$)$.}
\end{figure}

Following the order of investigation in the Coulomb phase, we start with the
correlator between gauge-fields with $1$ and $\sqrt{2}$ times the minimum
spatial momentum. The result is presented in Fig.\ 4.
With either procedure the two correlators seem to lie on top of each other,
indicating that the effective mass (\ref{meff}) is dominated by the $m^2$
contribution.
This confirms our expectation that there is no massless excitation in the
disordered phase, but the two graphs show different slopes so that the
resulting effective mass is a gauge-dependent object.
Moreover, it need not agree with what is usually called the photon mass, namely
the asymptotic mass in the $1^{+-}$ plaquette-plaquette correlator
\cite{BergCox}, which is a gauge-invariant quantity.
The gauge-dependence of the effective mass of the correlator (\ref{carb})
for $\be<\be_\mr{c}$ is also evident from Fig.\ 5 and the Table below.
Note finally that in the confined phase the Polyakov loop correlator is so
noisy that in practice no signal can be extracted.

\begin{figure}[t]
\vspace{-2mm}
\epsfig{file=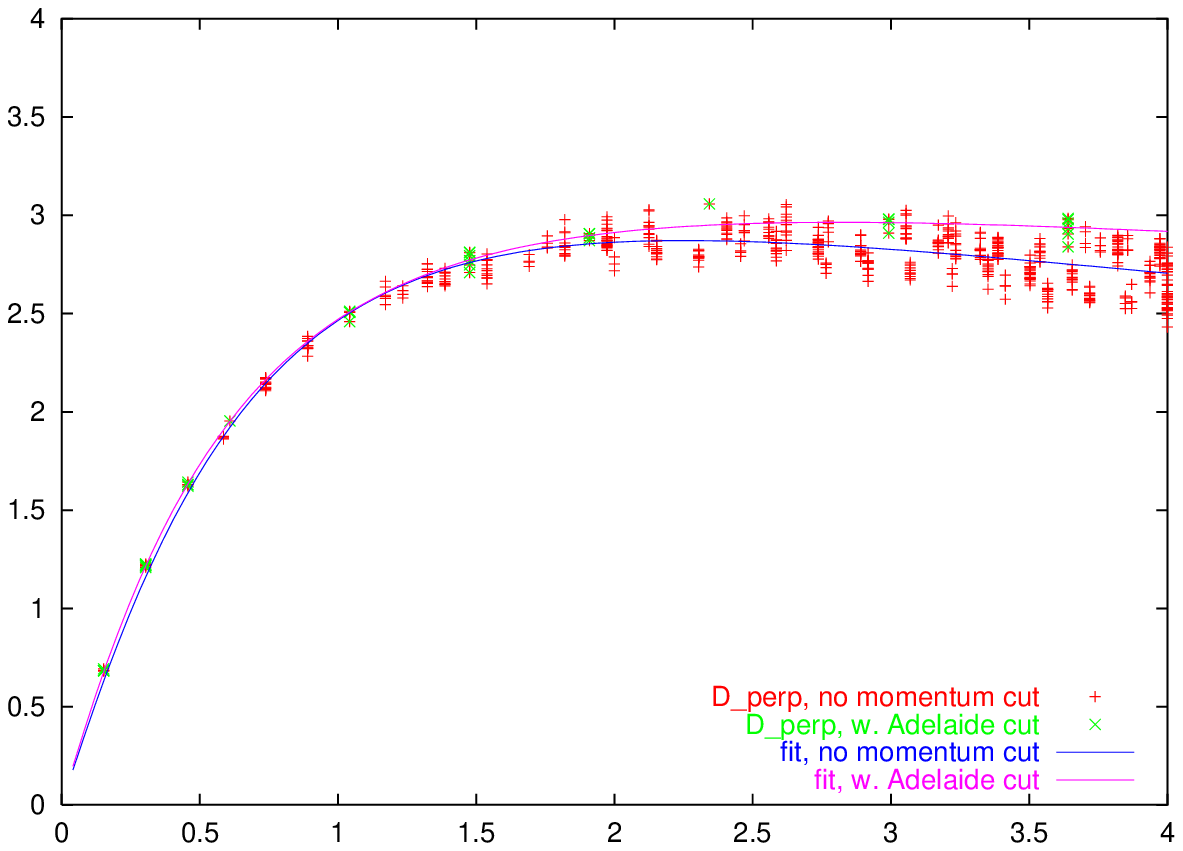,height=6cm,width=8.6cm,angle=0}
\hfill\hspace{-5mm}\hfill
\epsfig{file=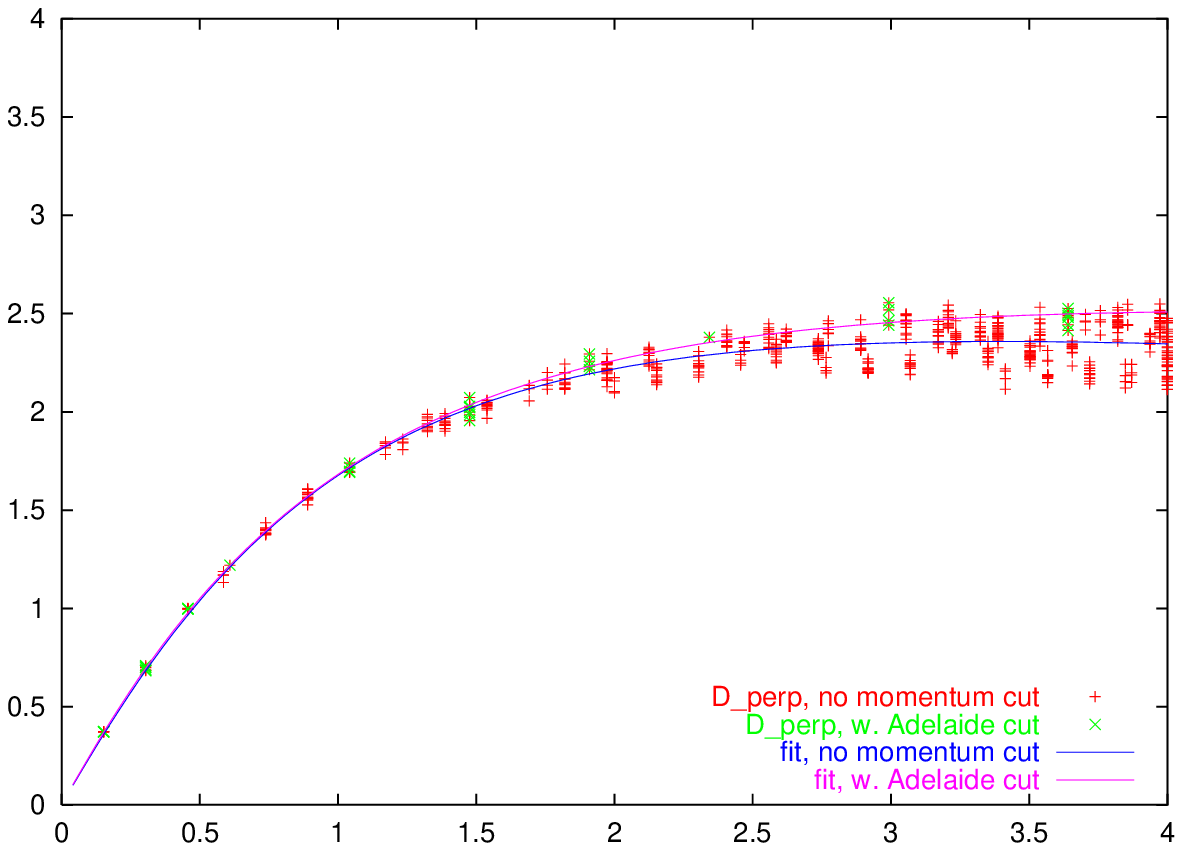,height=6cm,width=8.6cm,angle=0}
\\
\epsfig{file=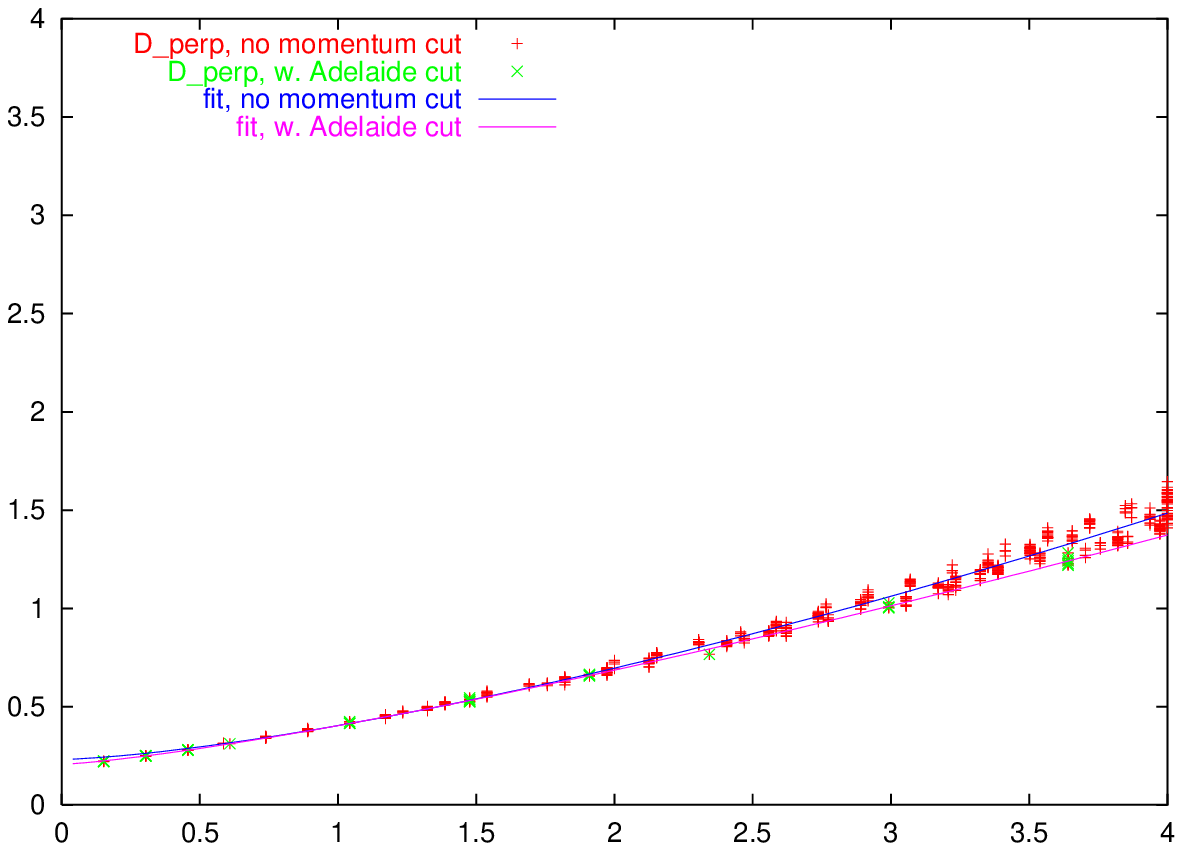,height=6cm,width=8.6cm,angle=0}
\hfill\hspace{-5mm}\hfill
\epsfig{file=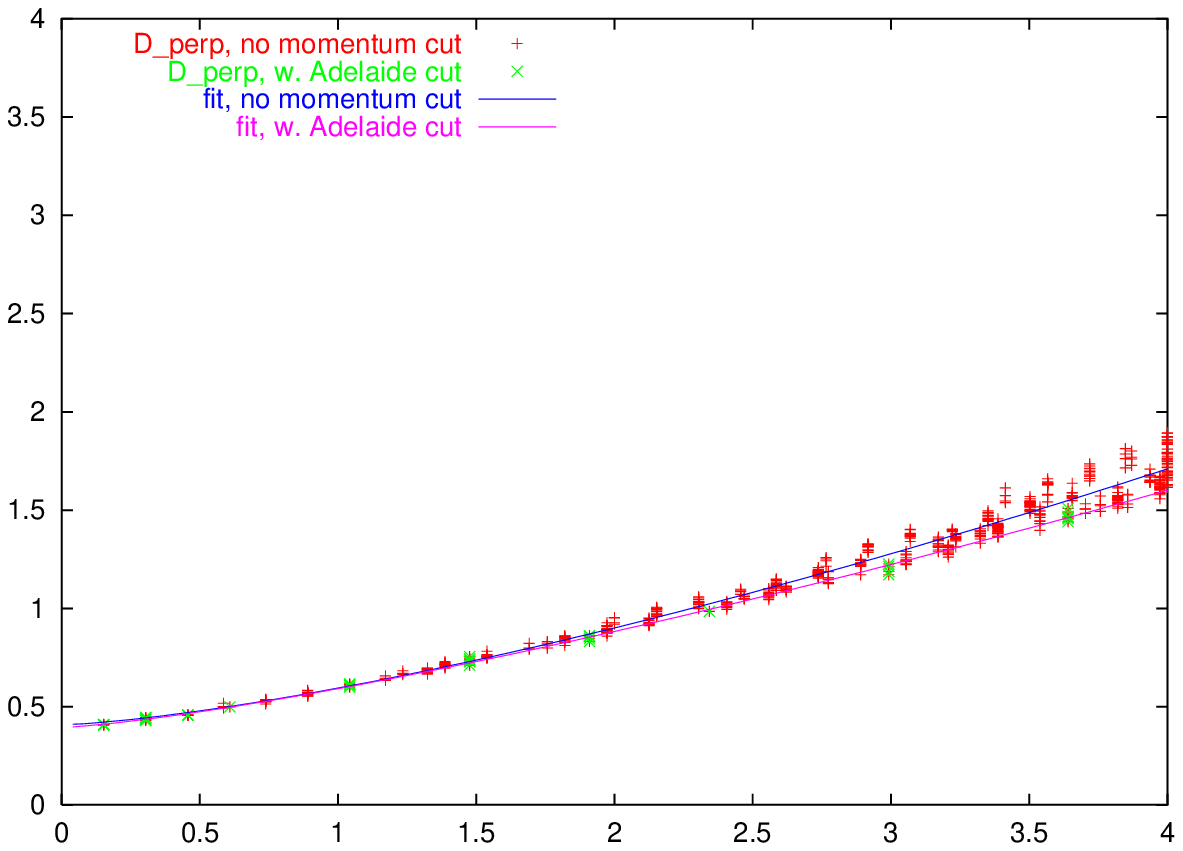,height=6cm,width=8.6cm,angle=0}
\vspace{-8mm}
\caption{\sl\small
Transverse component of the correlator (\ref{carb}) at $\beta\!=\!0.95$ on a
$16^4$ lattice, after modified iterative $($left$)$ and Laplacian $($right$)$
Landau gauge fixing. We plot $\hat p^2 D_\perp(\hat p^2)$ vs.\ $\hat p^2$
$($top$)$ and $1/D_\perp(\hat p^2)$ vs.\ $\hat p^2$ $($bottom$)$.
Both gauges yield a propagator which is not perfectly rotationally invariant,
since there is a certain amount of scatter among the data for a given
$\hat p^2$ $(+)$. After the cylindrical momentum cut $(\times)$ the scatter is
much reduced. Note that the intercept, in the bottom line, of the correlator
with the y-axis is larger in the Laplacian gauge than in the modified iterative
case.}
\end{figure}

In order to investigate the difference between the two fixing procedures, we go
through the same steps as before: We consider the complete correlator
(\ref{carb}) involving arbitrary four-momentum $\hat p_\mu$.
The result for the confined phase is shown in Fig.\ 5.
We plot $\hat p^2 D_\perp(\hat p^2)$ vs.\ $\hat p^2$ (top) and
$1/D_\perp(\hat p^2)$ vs.\ $\hat p^2$ (bottom) -- both after modified iterative
(left) and Laplacian (right) fixing.
Unlike in the Coulomb phase, rotational invariance is not a problem, i.e.\ the
data for a given $\hat p^2$ collapse on a single line in either gauge.
From a qualitative point of view it appears that in the confined phase
(\ref{carb}) is less sensitive to the details of the fixing procedure.
On a quantitative level, however, the agreement is substantially less
impressive.
Again, we fit the correlators to the ansatz (\ref{propansatz}), which now seems
much more adequate than in the ordered phase.
The resulting parameters (cf.\ Table below) indicate that the confined phase is
characterized by a distinctly {\em nonzero mass parameter\/} $m$ and a
{\em positive anomalous dimension\/} $\al$, but the mass in LLG is
substantially larger than that in the modified ILG.

\begin{table}[!b]
\begin{center}
\begin{tabular}{|c|ccc|ccc|}
\hline
{$\mb{\be\!=\!0.95}$} & {}&$\!\!\!\!${\bf modif.\ ILG}$\!\!\!\!$&{} &
                        {}&{\bf LLG}&{}
\\
{(confined)}       & $Z$&$m$&$\al$ & $Z$&$m$&$\al$
\\
\hline
{all momenta}      & 5.25$\pm$0.04&1.10$\pm$0.02&0.41$\pm$0.01 &
                     4.29$\pm$0.05&1.33$\pm$0.02&0.40$\pm$0.02
\\
{\bf cylind.\ cut} & 5.01$\pm$0.02&1.02$\pm$0.01&0.28$\pm$0.01 &
                     4.29$\pm$0.02&1.30$\pm$0.01&0.29$\pm$0.01
\\
\hline
\end{tabular}
\vspace{-5mm}
\end{center}
\caption{\sl\small Coefficients of the fits of $\hat p^2D_\perp(\hat p^2)$ vs.\
$\hat p^2$ to $(\ref{propansatz})$ in the confined phase -- see Fig.\ 5.
The two lines represent the result without and with cylindrical momentum cut
$($see text$)$. Note that the main uncertainty comes from the choice of
the fitting method, not from statistics.}
\end{table}

Again we check on the transversality.
The modified iterative correlator is (almost perfectly) transverse, the
Laplacian one is not.
However, the non-transversality in LLG is seriously reduced compared to what we
found in the Coulomb phase and rotational invariance is well respected in the
transverse piece, too (see Fig.\ 3).
In passing we note that these findings are consistent with the hypothesis
(which will be discussed below) that the Laplacian Landau gauge lacks the
ability to fully remove the DDS in the Coulomb phase -- rotational invariance
is restored in the confined phase, since the DDS percolate at the transition
\cite{percolation}.


Finally, we check whether any difference between the two procedures may be seen
in the distribution of the link angles, and here the answer is affirmative:
In the confined phase, the distribution after Laplacian fixing 
deviates much more visibly from a Gaussian than that after iterative fixing.


\section{Discussion and Conclusion}

Having established that on either side of the phase transition in the compact
$U(1)$ theory the correlator (\ref{carb}) in the modified iterative Landau
gauge (ILG) differs from that in the Laplacian Landau gauge (LLG), it is
natural to ask which are the specific properties of the gauge fixing procedures
which cause these differences.

From the discussion of the gauge fixing procedures in sect.\ 2, it is clear
that double Dirac sheets and zero-momentum modes represent the primary
candidates for gauge artefacts which might spoil the properties of a gauge
(cf.\ \cite{Bogolubsky:1999cb}).
Hence, a natural thing to try is to omit the DDS suppression in the iterative
procedure while keeping the step which reduces the ZMM background; this is what
we have introduced as our ``standard'' ILG.
The result of such a gauge fixing is presented in Figs.\ 6 and 7.
In the Coulomb phase, the correlators with minimum or next-to-minimum spatial
momentum (l.h.s.\ of Fig.\ 6) are flatter than those after modified ILG and LLG
(cf.\ Fig.\ 1), indicating a {\em negative\/} $m^2$.
This means that the DDS tend to decrease the effective mass (\ref{meff}).
In the confined phase the ``standard'' iterative correlators essentially agree
with those in the modified iterative gauge.
A more detailed picture is gotten by considering arbitrary four-momenta, as
displayed in Fig.\ 7.
The resulting longitudinal piece is essentially zero, hence only the transverse
part is shown.
The startling observation is that the ``standard'' ILG propagator shows the
same amount of scatter and the same sort of IR enhancement in the Coulomb phase
that we have seen in the case of the Laplacian correlator.
On the other hand, in the confined phase the ``standard'' iterative propagator
almost coincides with its modified counterpart.
This is expected, since the DDS percolate at the phase transition
\cite{percolation}; hence the efforts of the modified ILG to remove them are
futile.

\begin{figure}[t]
\vspace{-2mm}
\epsfig{file=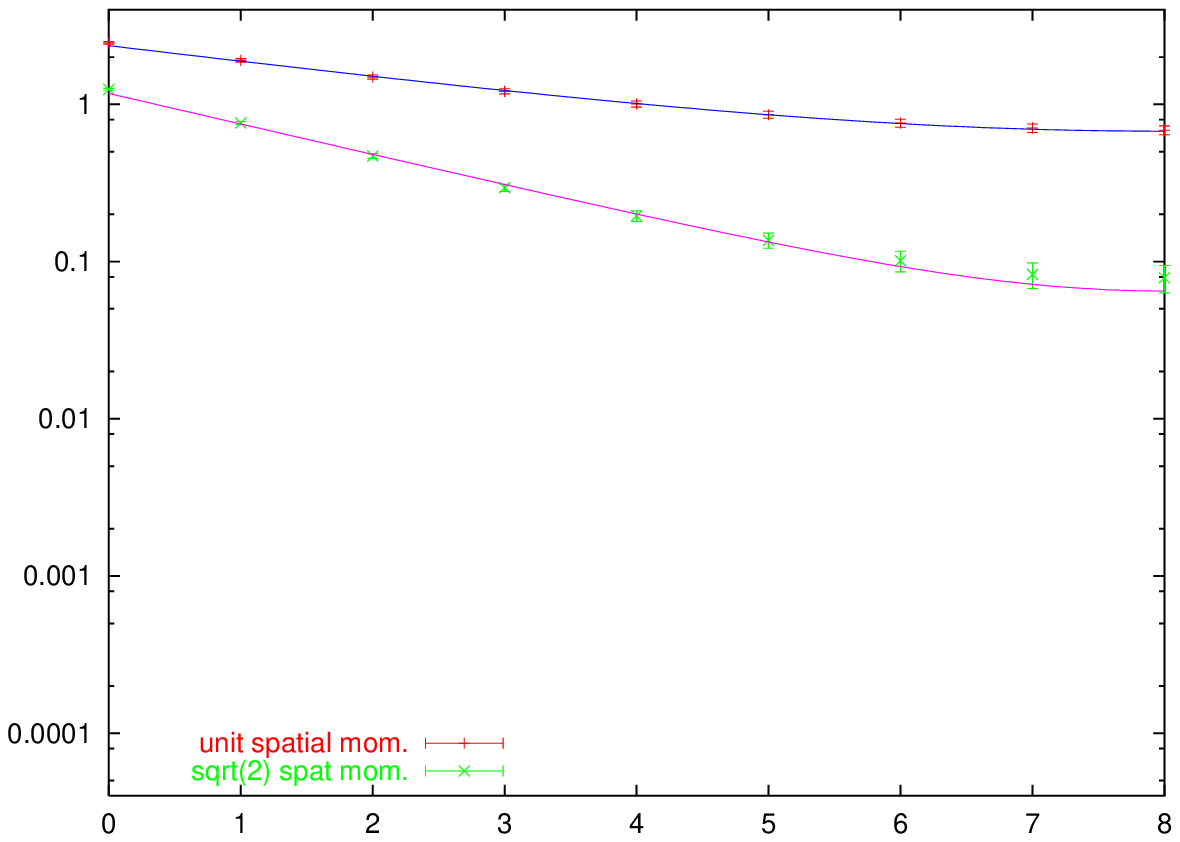,height=6cm,width=8.6cm,angle=0}
\hfill\hspace{-5mm}\hfill
\epsfig{file=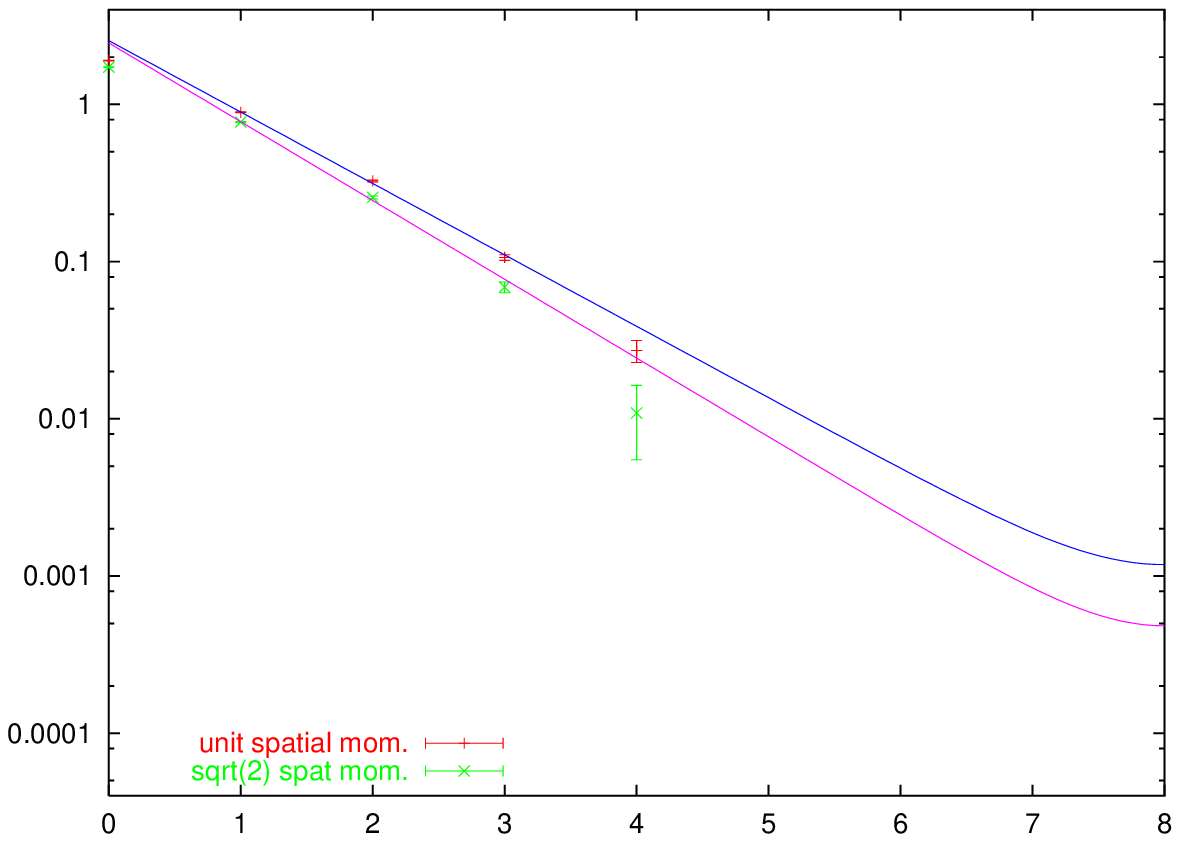,height=6cm,width=8.6cm,angle=0}
\vspace{-8mm}
\caption{\sl\small
Field (\ref{photdef}) correlator with with once $(+)$ or $\sqrt{2}$ times the
minimum spatial momentum in the ``standard'' iterative Landau gauge at
$\be\!=\!1.05$ $($l.h.s.$)$ and at $\be\!=\!0.95$ $($r.h.s$)$.
The propagator in the Coulomb phase is flatter than its modified iterative or
Laplacian counterparts at the same $\be$ value (cf.\ Fig.\ 1), while in the
confined phase it is in good agreement with its modified ILG counterpart but
not with the Laplacian one (cf.\ Fig.\ 4).}
\end{figure}

\begin{figure}[t]
\vspace{-2mm}
\epsfig{file=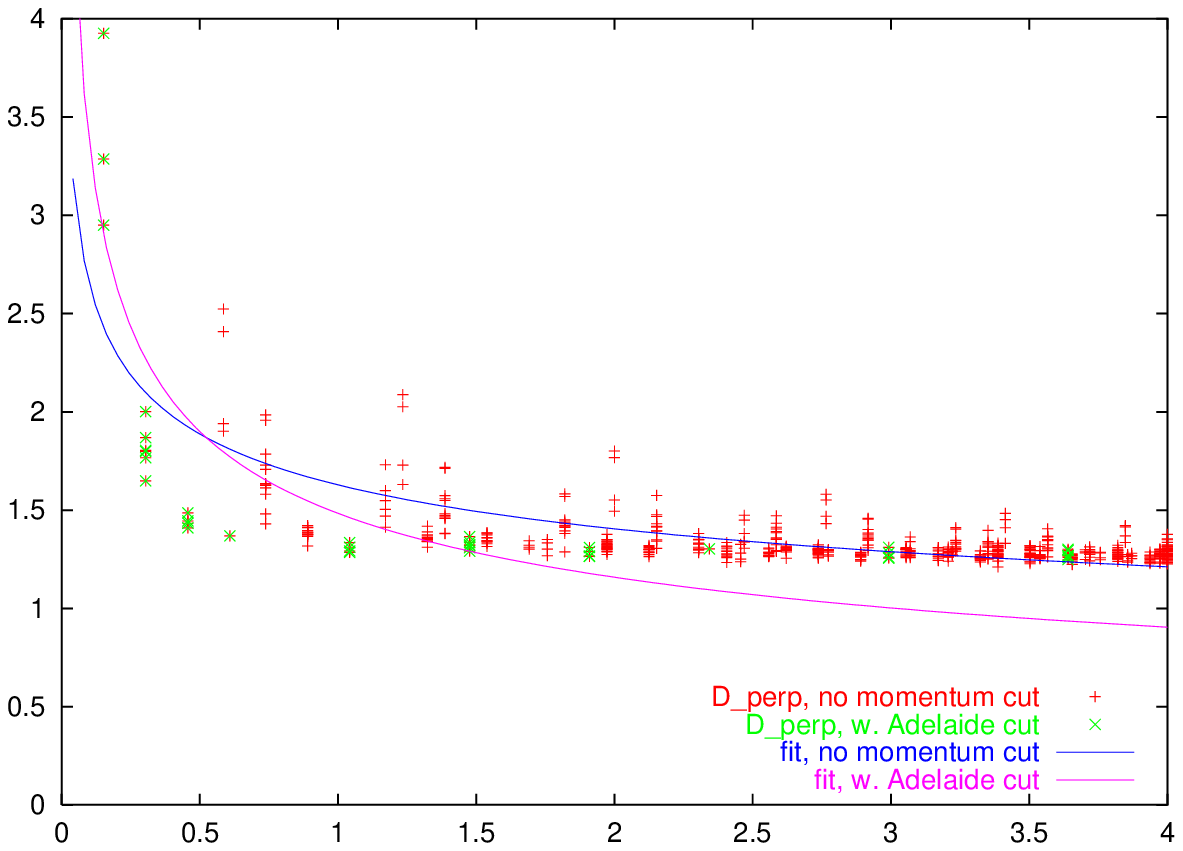,height=6cm,width=8.6cm,angle=0}
\hfill\hspace{-5mm}\hfill
\epsfig{file=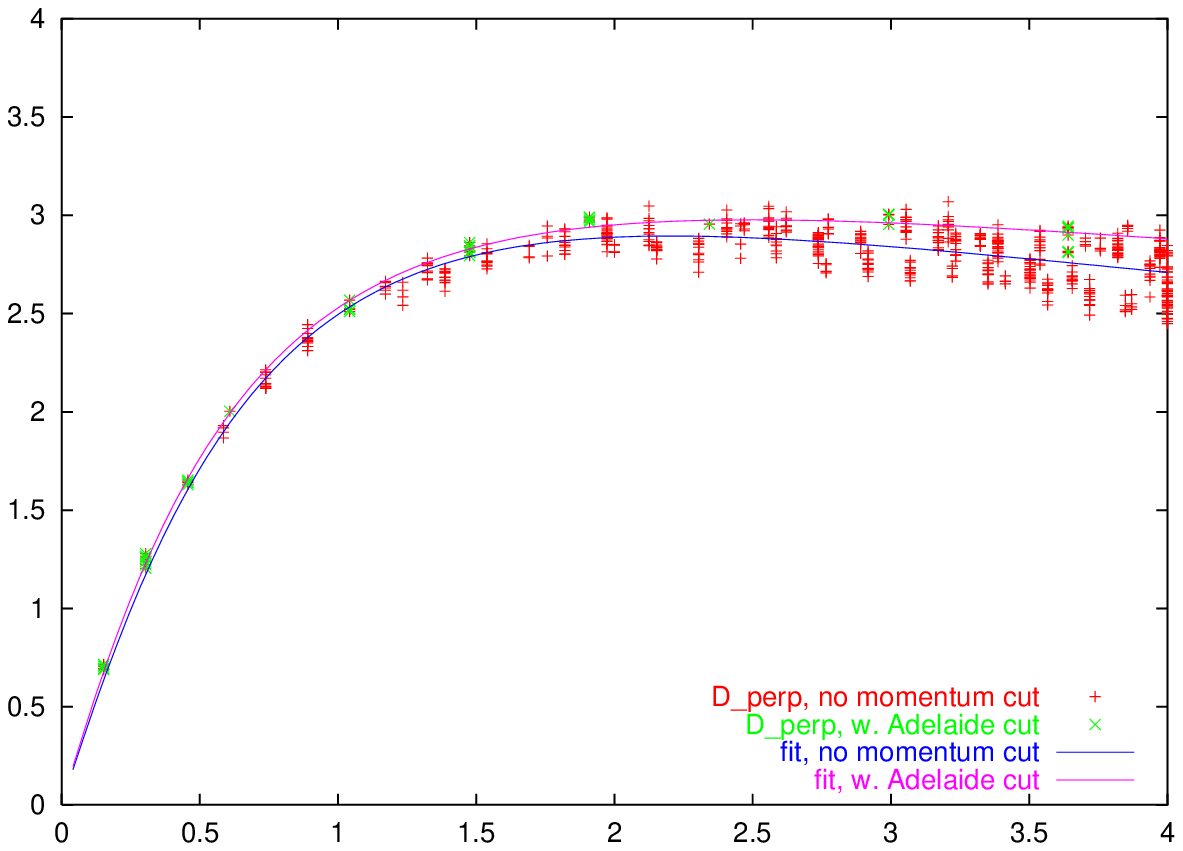,height=6cm,width=8.6cm,angle=0}
\\
\epsfig{file=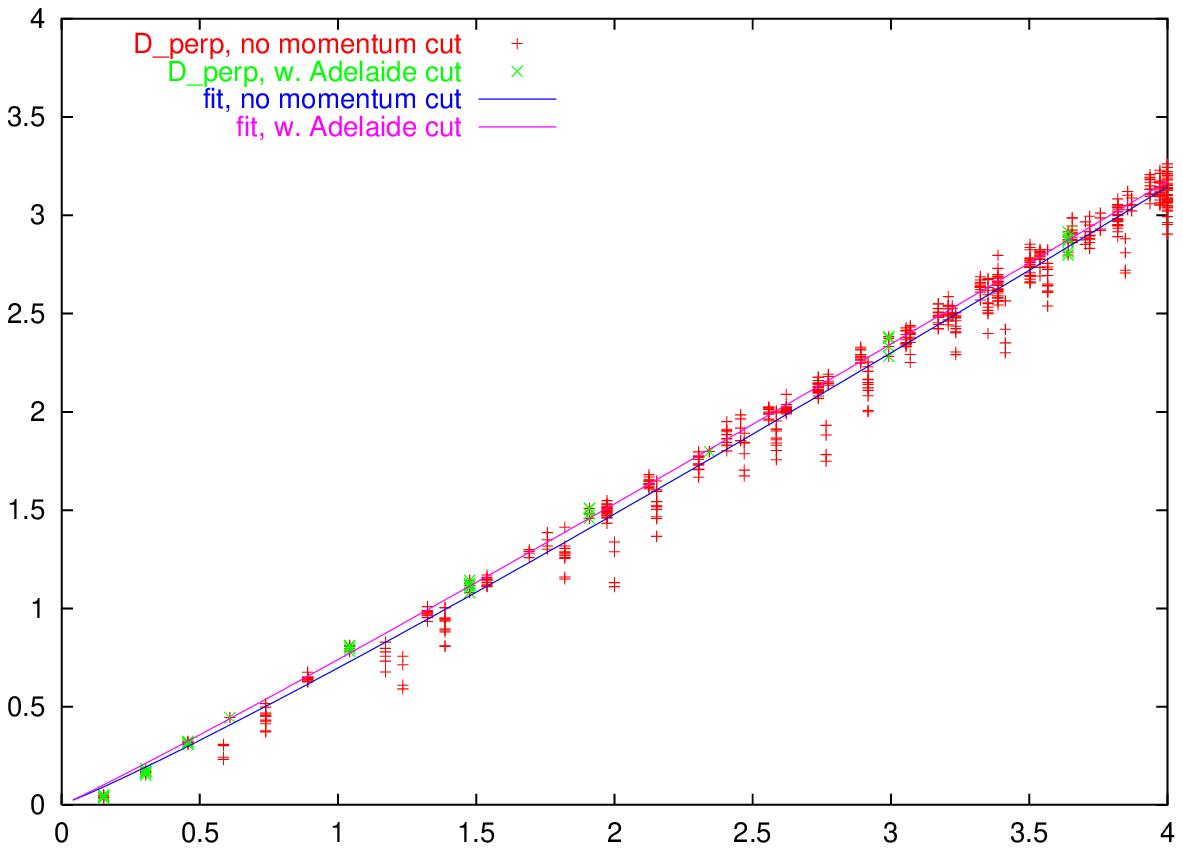,height=6cm,width=8.6cm,angle=0}
\hfill\hspace{-5mm}\hfill
\epsfig{file=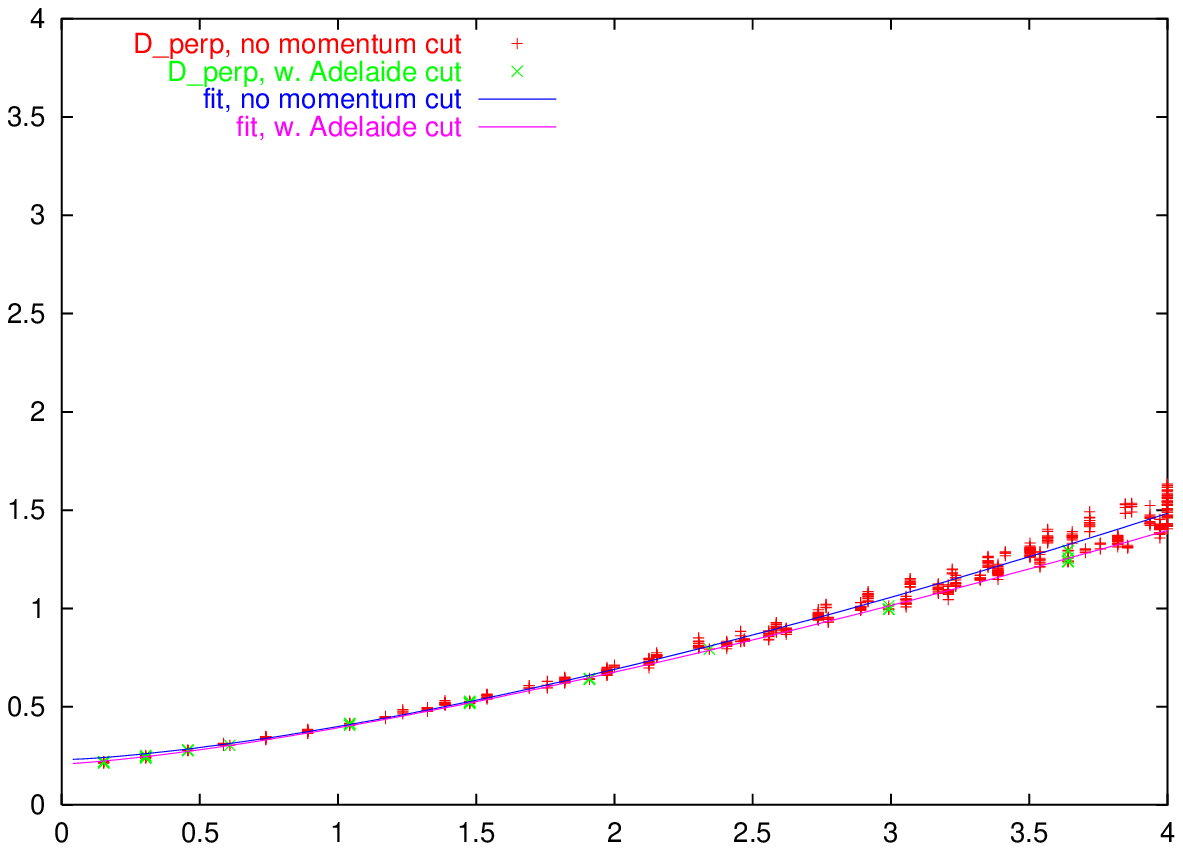,height=6cm,width=8.6cm,angle=0}
\vspace{-3mm}
\caption{\sl\small
Photon correlator in the ``standard'' iterative prescription at $\beta=1.05$
$($left$)$ and at $\beta=0.95$ $($right$)$. Each time the graphs for
$\hat p^2 D_\perp(\hat p^2)$ vs.\ $\hat p^2$ $($top$)$ and $1/D_\perp(\hat p^2)$
vs.\ $\hat p^2$ $($bottom$)$ are shown. In the Coulomb phase the remaining DDS
cause a massive IR enhancement which is absent in the modified ILG counterpart
but equally pronounced in LLG (cf.\ Fig.\ 2). In the confined phase the result
agrees with the modified ILG correlator (cf.\ Fig.\ 5).}
\end{figure}

In an attempt to further substantiate the obvious suspicion that remnant DDS
are responsible for the abnormal IR behaviour of the Laplacian propagator in
the Coulomb phase and to shed light on the properties of the Laplacian gauge in
the confined phase, we produce a scatter plot relating the value of the
functional (\ref{func}) to the remaining number of Dirac plaquettes in the
gauge fixed configuration. The result is shown in Fig.\ 8.

\begin{figure}[t]
\vspace{-2mm}
\epsfig{file=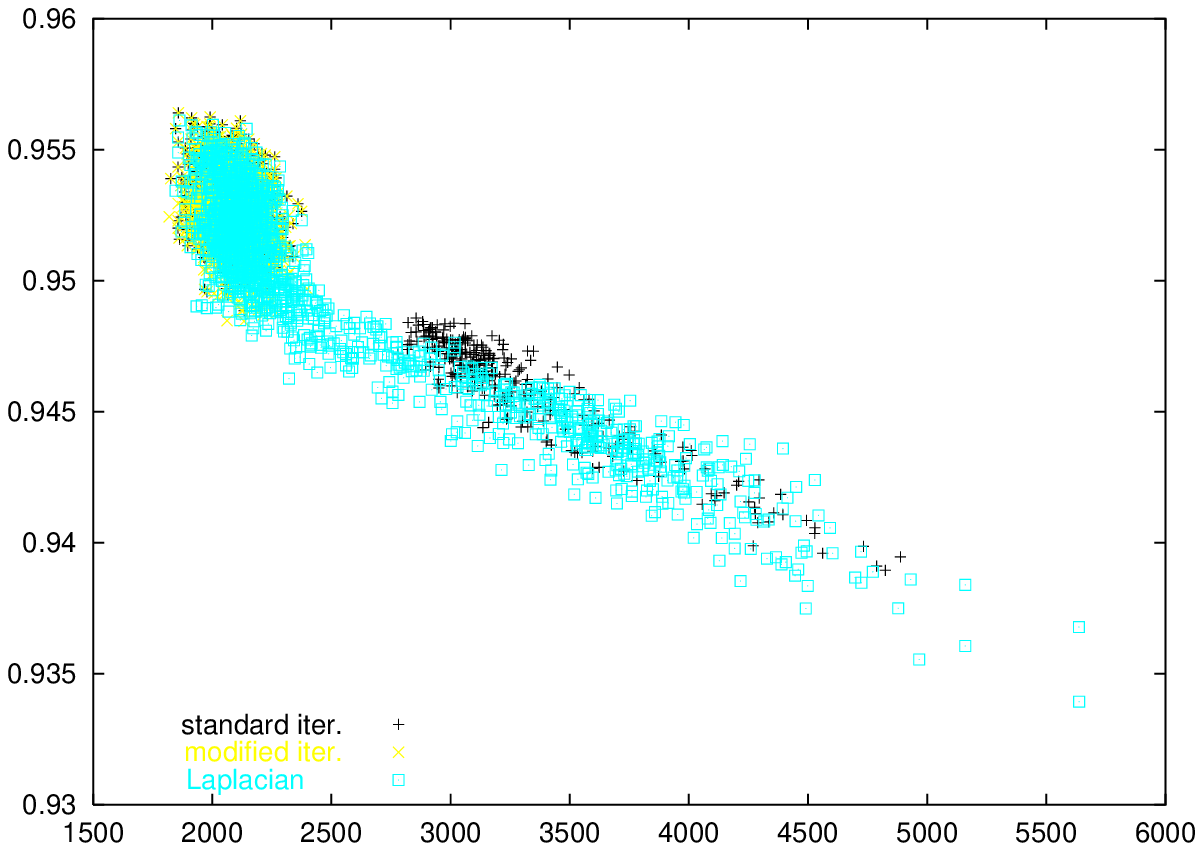,height=6cm,width=8.6cm}
\hfill\hspace{-5mm}\hfill
\epsfig{file=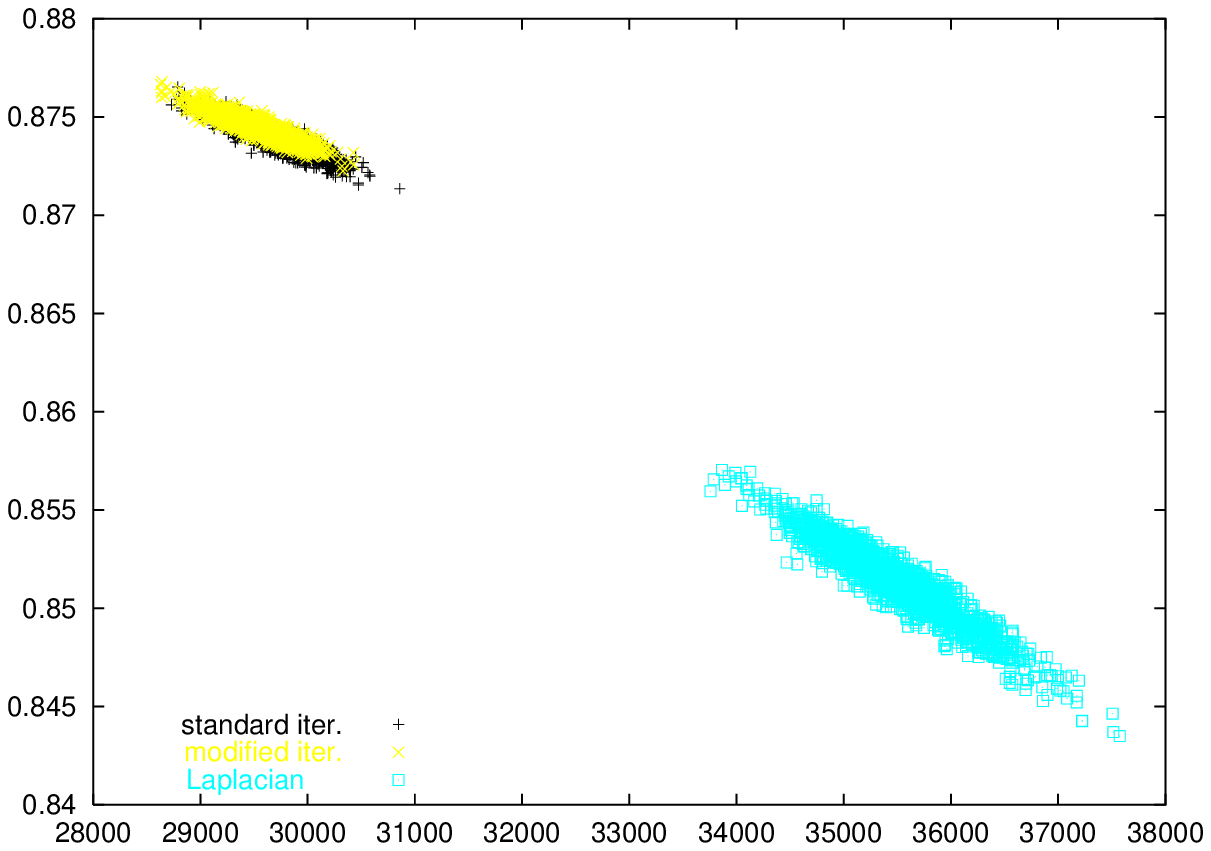,height=6cm,width=8.6cm}
\vspace{-8mm}
\caption{\sl\small
Correlation between the final value of the functional $(\ref{func})$ and
the number of Dirac plaquettes in the fixed configuration in the Coulomb phase
$($l.h.s.$)$ and in the confined phase $($r.h.s.$)$.}
\end{figure}

In the Coulomb phase the modified ILG always reaches the highest value of
(\ref{func}), among the three gauges considered.
The ``standard'' ILG reaches (almost) the same extremum on the majority of
configurations, but there is a considerable number of backgrounds on which it
falls short quite drastically, and at the same time the total number of Dirac
plaquettes is substantially increased.
This correlation is paralleled in LLG, but there the amount by which it falls
short (in either respect) has a continuous spectrum.
In other words, the ``standard'' ILG differs from the LLG by the gap in the
functional value which separates ``success'' from ``failure''.
Hence it looks like only a limited number of configurations causes
(simultaneously for the Laplacian and the ``standard'' iterative procedure)
a problem both w.r.t.\ the gauge functional and the number of Dirac plaquettes.
We checked that these are, in most cases, the same configurations for both
gauges.
On our lattice the DDS condition (\ref{DDScond1}) reads
$2\!\cdot\!16^2\!=\!512$ per plane.
With a typical number of $O(2000)$ Dirac plaquettes at $\be\!=\!1.05$ in a
modified ILG configuration, one would expect a configuration with a DDS sheet
to have more than $O(2500)$ Dirac plaquettes.
This is precisely where the gap in the number of Dirac plaquettes for
``standard'' iterative fixing ends.
Since the Laplacian fixing shows a continuous excess in the number of Dirac
plaquettes, this correlation analysis cannot prove that DDS present a problem
to the LLG procedure.
However, the condition (\ref{DDScond1}) being met, together with our previous
observation that rotational invariance is seriously broken (in particular for
small momenta, cf.\ Fig.\ 3) is strong circumstantial evidence.

In the confined phase things look much simpler:
Again, the modified ILG wins the competition for the extremization of
the gauge functional (\ref{func}).
The ``standard'' ILG is in second rank, but it is very close -- both w.r.t.\
the functional value and the number of remnant Dirac plaquettes.
The LLG is far off -- again w.r.t.\ both criteria.
The only surprise is by how much it misses the extremum and by how many Dirac
plaquettes it deviates from the modified ILG.
Again this establishes a link to our previous finding that the LLG respects
rotational invariance in the confined phase:
If unresolved DDS in the Coulomb phase are responsible for the shortcomings
of the Laplacian gauge, then one would expect, since the DDS percolate
at the transition \cite{percolation}, that rotational invariance
gets restored in the confined phase while the excess of Dirac plaquettes
remains -- and this is precisely what we see.

While these preliminary observations clearly show a demand for future research,
we would like to close with a few general remarks:
\begin{itemize}
\item
At first sight, it looks like some of our correlators are quite ``fuzzy''.
Unlike many other groups we have refrained from averaging over momenta with the
same $\hat p^2$; if one would take the average $D_\perp(\hat p^2)$, then all
propagators would become thin lines.
However, as is obvious from the upper right plot in Fig.\ 2, such a procedure
would only reveal a {\em slight\/} IR enhancement,
and the information about the lack of rotational invariance would be lost.
Hence we believe it is important to retain the full information about the
spread of correlator points associated with a given $\hat p^2$.
\item
It is not clear whether the qualitative aspects of the propagators we have
studied on a $16^4$ grid stay invariant under a change of the lattice size.
%
%
On one hand the ability of the ILG to remove DDS through repeated attempts
will degrade drastically on larger lattices; presumably the cut-off on the
outer loop (as discussed in sect.\ 2) should be increased exponentially with
the volume.
On the other hand DDS on a larger lattice involve more Dirac plaquettes and
become less frequent by purely entropic suppression.
Which effect wins is unclear.
Preliminary studies on a $32^4$ grid indicate that the Laplacian correlator
might be much better behaved on that lattice size -- unfortunately the
available statistics has prevented us from reaching any conclusion.
Still, our $16^4$ lattice is much larger than those which have been used in
previous investigations of the modified ILG \cite{Bogolubsky:1999cb}.
\item
After dealing with so many details of which quantity correlates with which
other one, we wish to emphasize a simple point:
We have presented evidence that all three gauges considered yield {\em smooth
configurations in either phase\/}.
At $\be\!=\!1.05$ we see no differences in the distribution of link angles
at all, whereas at $\be\!=\!0.95$ the configurations tend to be slightly
smoother with both types of iterative fixing than in the Laplacian gauge.
Notably, in spite of the distribution of link angles being almost
indistinguishable, the propagators (\ref{carb}) on these ensembles may be
strikingly different.
\item
The lack of several gauge fixing procedures to reproduce the known perturbative
behaviour in the Coulomb phase of the compact $U(1)$ theory raises concerns
about the validity of results for $SU(2)$ or $SU(3)$ in either standard
iterative or Laplacian gauge.
If already a $U(1)$ subgroup may cause a problem, one does not feel inclined to
blindly trust the result of any fixing procedure in the non-Abelian case.
The mass which we have extracted from the gauge-fixed field correlator is
not gauge-invariant (see Table~1).
This feature is likely to persist in the non-Abelian case below $T_c$.
Therefore, the quantitative meaning of gauge-fixed gluon mass measurements
should be considered with caution.
It will be interesting to see whether our observation that the deconfined phase
is more sensitive to the details of the fixing procedure carries over to the
non-Abelian case, too.
\end{itemize}


\noindent
{\bf Acknowledgements:}
Out of the vast body of institutional addresses, one deserves to be highlighted:
The Institute for Nuclear Theory is where the two of us met during the program
``Lattice QCD and Hadron Phenomenology'' (INT-01-3) and the stimulating
atmosphere genuine to this place spurred us on to complete the current project.
One of us (S.D.) acknowledges useful discussions with Urs Heller and
Karl Jansen.


\end{document}